\documentclass[a4paper,fleqn,usenatbib]{mnras} 

\usepackage{newtxtext,newtxmath}

\usepackage[T1]{fontenc}
\usepackage{ae,aecompl}

\usepackage{graphicx}	
\usepackage{amsmath}	
\usepackage{amssymb}	
\usepackage{tikz}
\usetikzlibrary{shapes,arrows,positioning,plotmarks,patterns,decorations.pathmorphing,decorations.markings,intersections}
\usepackage{pgf}
\usepackage{pgfplots}
\usepackage{units}

\title[ELT distortions on astrometric observations]{The impact of ELT distortions and instabilities on future astrometric observations}

\author[G. Rodeghiero et al.]{
G. Rodeghiero,$^{1}$\thanks{E-mail: rodeghiero@mpia.de}
J.-U., Pott,$^{1}$
C. Arcidiacono,$^{2}$
D. Massari,$^{3}$
M. Gl\"uck,$^{4,1}$,
\newauthor
H. Riechert,$^{1}$
E. Gendron$^{5}$
\\
$^{1}$Max Planck Institut f\"ur Astronomie, K\"onigsthul 17, D-69117, Heidelberg, Germany\\
$^{2}$INAF Osservatorio Astrofisico e scienza dello Spazio di Bologna (OAS), Via Gobetti 93/3, 40129 Bologna, Italy\\
$^{3}$Kapteyn Astronomical Institute, University of Groningen, NL-9747 AD Groningen, Netherlands\\
$^{4}$Universit\"at Stuttgart, Institut f\"ur Systemdynamik, Waldburgstr. 19, 70563 Stuttgart, Germany\\
$^{5}$LESIA, Observatoire de Paris, CNRS, UPMC, Universit\'e Paris Diderot, 5, place Jules Janssen, 92190 Meudon, France\\
}

\date{Accepted XXX. Received YYY; in original form ZZZ}

\pubyear{2017}

\begin{document}
\label{firstpage}
\pagerange{\pageref{firstpage}--\pageref{lastpage}}
\maketitle

\begin{abstract}
The paper discusses an assessment study about the impact of the distortions on the astrometric observations with the Extremely Large Telescope originated from the optics positioning errors and telescope instabilities. Optical simulations combined with Monte Carlo approach reproducing typical inferred opto-mechanical and dynamical instabilities, show RMS distortions between $\sim$ 0.1-5 mas over 1 arcmin field of view. Over minutes timescales the plate scale variations from ELT-M2 caused by wind disturbances and gravity flexures and the field rotation from ELT-M4-M5 induce distortions and PSF jitter at the edge of 1 arcmin FoV (radius 35 arcsec) up to $\sim$ 5 mas comparable to the diffraction-limited PSF size $FWHM_H = 8.5$ mas. The RMS  distortions inherent to the ELT design are confined to the 1$^{st}$-3$^{rd}$ order and reduce to an astrometric RMS residual post fit of $\sim$ 10-20 $\mu as$ for higher order terms. In this paper, we study which calibration effort has to be undertaken to reach an astrometric stability close to this level of higher order residuals. The amplitude and timescales of the assumed telescope tolerances indicate the need for frequent on-sky calibrations and MCAO stabilization of the plate scale to enable astrometric observations with ELT at the level of $\leq 50 \mu as$, which is one of the core science missions for the ELT / MICADO instrument.

\end{abstract}

\begin{keywords}
Distortion -- Astrometry -- Centroid jitter -- Positioning Errors -- Tolerances
\end{keywords}



\section*{Introduction}
\label{intro}

\label{onecolumn}
Astrometry with large telescopes is one of the most challenging observation modes in modern, ground-based astronomy. The new generation of extremely large telescopes, the Extremely Large Telescope (ELT, 39 m) \cite{Tamai2014}, the Thirty Meter Telescope (TMT, 30 m) \cite{Simard2013}, the Giant Magellan Telescope (GMT, 24.5 m) \cite{McCarthy2016}, with their huge primary mirrors, will boost both the sensitivity and the spatial resolution, paving the road to a new level of high-resolution astronomy. The enhanced performances of this new generation of telescopes will lead to $\sim$ 50 $\mu$as astrometry for ground-based near-infrared (NIR) imagers over significant field sizes. Instruments like MICADO \cite{Davies2016} for ELT and IRIS \cite{Larkin2016} for TMT, aim to deliver  $\simeq$ 50 $\mu$as post-fit differential astrometry within a single epoch, leading to a jump with respect to the astrometric noise floor of the current instruments as NIRC2 at Keck and WFC3 on HST $\sim$ 0.15 mas \cite{Lu2014}, and GeMS $\sim$ 0.4 mas \cite{Neichel2014}. 
The average, achievable precision for centroiding the high-resolution core of a PSF is approximately:

\begin{equation}\label{eq1}
\sigma_{x,y} \approx \frac{\sigma_{PSF}}{SNR}
\end{equation}


with $\sigma_{PSF}$ being standard deviation of the Gaussian fit of the PSF core and SNR the signal to noise ratio of the observation. 



Strictly applying \ref{eq1} in the passage from 8m class telescope to the ELT leads to $\sigma_{ELT}\geq  \sigma_{8m}/125$, due to the five times smaller diffraction limited FWHM and an SNR gain factor 5 or 25  for photon-noise or background-noise limited astrometry. The sensitivity gain factor scales with $\sim D^3$. The new generation of ELTs aims for $\sigma_{centroid}/\sigma_{PSF} \sim 1/100$, in $H$ band this means  $\sigma_{centroid} \sim$ 40 $\mu$as. However, a larger diameter $D$ implies also an increase of the overall telescope size and mass that results in a more complex telescope structure and control strategy. Defining $S$ as representative scale size parameter of the telescope (e.g. $D$), the deflections due to gravity by self weight scale as $\delta_g \sim S^2$ and the mass of the structure as $\sim S^3$ \cite{Nelson1999}. Practical experience from Keck and simulation models for ELT assess a slightly softer dependence $M\sim kS^3$ with $k \sim 0.1-0.25$. So bigger telescopes translates also in bigger instabilities, e.g. $\delta_g(ELT) \sim 25\delta_g(8m)$, that challenge the ultimate astrometric precision of the observations. Instruments like MICADO and IRIS aim to achieve $\sim$ 50$\mu$as differential astrometry, i.e.  $\sigma_{ELT} =  \sigma_{8m}/5$, that seems to be achievable with a careful instrument design and calibration procedure. This precision is affected by a myriad of other systematics of different nature: instrumental, atmospheric and astronomical (\cite{Trippe2010}, \cite{Schock2014}). In this paper we concentrate only on the instrumental errors generated by the telescope instabilities and distortions, trying to model the behaviour of the ELT in typical operation scenarios. The approach we follow consists in injecting into the ELT nominal design some expected positioning errors of the optics to assess the impact of the latter on the distortion pattern variations and the residual astrometric errors.  
The paper after a first introduction of the telescope prescription data (Section \ref{s1}), discusses the simulation method used to assess the distortions on astrometry (Section \ref{s2}) and the results of the sensitivity study on the different ELT optics (Section \ref{s3}). Sections \ref{s4}, \ref{s5}, \ref{s6} report the results of three specific cases of study about the impact of M2 specific instabilities and the field rotation effects induced by M5.

\section{The ELT telescope}
\label{s1}

The Extremely Large Telescope (ELT) is the major ground-based observatory in the world under development at ESO and targeted for first light by the end of the next decade. The telescope optical configuration is based on an innovative 5 mirrors solution: a three-mirror on-axis anastigmat plus two fold mirrors with AO capabilities to correct high order atmospheric turbulence and tip-tilt jitter. The ELT is sensitive between $\lambda$ = 0.5-20 $\mu$m over a 10 arcmin Field of View (FoV) and it serves two Nasmyth foci where the instruments are accommodate at the sides of the rotatable telescope structure. As reported in Table \ref{tabsys}, the focal ratio of the telescope is $F/\#\sim 17.75$ and the plate scale $PS \sim 0.3$"/mm.

\begin{table}
\begin{center}
\begin{tabular}{p{5cm}*{1}{c}} \hline
\textbf{Telescope parameters} & \textbf{Size}\\
\hline 
\bfseries Effective focal length & 684021.6 mm\\
\bfseries Back focal distance & 27200 mm\\
\bfseries Working F/\# & 17.75\\
\bfseries Field of View & 0.08333333$^{\circ}$\\
\bfseries Obscuration ratio & $\sim$ 28.4\%\\
\bfseries Plate scale & 0.3016 ''/mm\\
\bfseries Angular magnification & 18.0635\\
\hline
\end{tabular}
\caption{Main ELT optical design parameters.\label{tabsys}}
 \end{center}
 \end{table}

\begin{figure*}
\centering\includegraphics[width=0.8\linewidth]{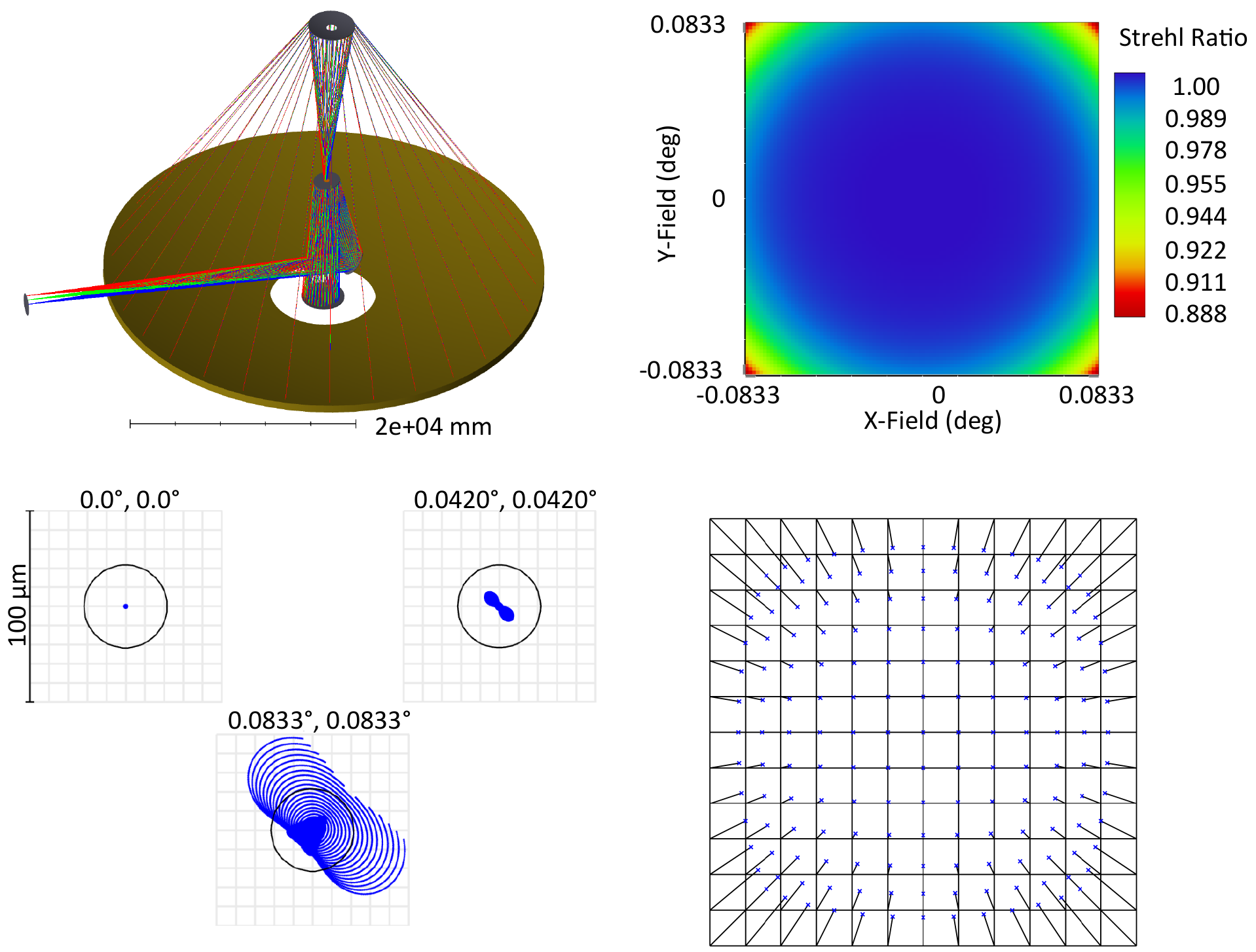}
\caption{\textbf{Top left}: 3D view of the ELT optics showing its three mirror anastigmat configuration plus M4 deformable mirror and M5 fast tip-tilt corrector redirecting the beam at the Nasmyth platform; \textbf{Top right}: Strehl ratio map at the ELT focal plane of the full FoV; \textbf{Bottom left}: on axis and off axis (2.52',2.52') and (4.98',4.98')  ELT PSFs in comparison to the theoretical Airy disk (black circle), scale bar in micron; \textbf{Bottom right}: azimuthally symmetric distortion pattern at the ELT focal plane (scale x100), the maximum distortion at the corner of the FoV is $\sim$ 0.263 \%. \label{ima1}}
\end{figure*}

In the following subsections we report the main optical specifications of the five ELT mirrors as used in this paper. A summary of the main ELT optics specifications is collected in Table \ref{tabgeo} while Figure \ref{ima1} shows a 3-D view of the ELT, the nominal Strehl ratio map and the PSFs over the full FoV, and a map of the geometrical distortions of the telescope.

\subsection*{M1}
The primary mirror has an elliptical prolate profile with a diameter of $\sim$ 38 m and consists of 798 hexagonal segments, each  1.5 metres across and 50 mm thick, to be kept in phase by  edge sensors that measure the differential displacements between adjacent segments. An active optics mechanism controls piston, tip and tilt of each segment by means of three position actuators using the data of the edge sensors, which are calibrated periodically ($\sim$ 2 weeks) on sky \cite{Bonnet2011}. The warping harness of the actuators allows also to change/re-adjust the shape of the segment and to keep within the optical specifications the surface of the M1 against gravity, wind and temperature perturbations. 

\subsection*{M2}
The secondary mirror is a convex hyperbolic and aspheric meniscus made of Zerodur with a diameter of $\sim$ 4.1 m. As discussed in the next sections, M2 is a critical component of the ELT optical design, given its high optical sensitivity to positioning errors and its position in the telescope structure ($\sim$ 30 m away from M1). The weight of the M2 cell (optics+mechanics and harness) is $\sim$12 ton and the mirror is passively supported by an 18 point axial whiffletree. A warping harness system allows to correct the low order deformations of M2, but due to its high optical sensitivity, the cell needs to be repositioned periodically during the observations against the gravity flexures \cite{Mueller2014}. More details of M2 and the related problematics are discussed in Section \ref{s3} \ref{s4} and \ref{s5}.

\subsection*{M3}
The M3 is an aspheric concave mirror of $\sim$ 3.8 m diameter placed in the M1 central hole. The tertiary mirror allows refocusing and achieving a telescope variable focal length. The mirror control system requires the M3 to move instead of the secondary and therefore a flexible positioning system is used to shift the mirror in all six degrees of freedom; the M3 has a motion range of 250 mm in the direction of the M2, $\sim$ 200 mm allocated for focus change and $\pm$ 20 mm for compensating mirror prescription and integration errors \cite{Cayrel2012}. In addition, M3 is a thin meniscus with moderate active shape control forces.

\subsection*{M4}
The M4 is a 2.4-m flat deformable AO mirror that compensates for the wavefront errors due to misalignments of ELT optics, wind loads on the telescope structure and atmospheric disturbances \cite{Vernet2012}. The mirror is mounted on a positioning system providing a first stage large stroke low frequency mechanical tilt, a two dimensions  decentering degrees of freedom and a focus selector; it is inclined by 7.75$^\circ$ to allow switching between two symmetrical orientations of beam propagation towards the two telescope Nasmyth foci. The mirror is a thin Zerodur membrane of 1.95 mm thickness, segmented in six petals that are shaped by 5316 actuators using voice coil technology \cite{Biasi2016}. In median seeing conditions (0.85 arcsec) M4 shall provide a wavefront error smaller than 145 nm rms while in bad seeing conditions (1.1 arcsec) the residual fitting wavefront error shall be smaller than 180 nm rms \cite{Vernet2012}.

\subsection*{M5}
The M5, the last ELT mirror before the telescope focal plane, is a fast correcting optical element that provides tip-tilt corrections for the telescope dynamic pointing errors and the effect of atmospheric tip-tilt and wind disturbances. The M5 together with the M4 implements the pre-focal AO correction of the ELT \cite{Casalta2010}. The M5 is an elliptical mirror with dimensions of 2.4 m by 3 m. The full range on the mirror is around 60 arcsec in both axes with a control frequency range of 100 Hz.

\begin{table*}
\begin{center}
\begin{tabular}{p{1.6cm}*{5}{c}} \hline
\textbf{Surface} & \textbf{Diameter, mm} & \textbf{RoC, mm} & \textbf{F\#} & \textbf{Conic} & \textbf{Distance to, mm} \\
\hline 
\bfseries M1 & 38542 & 68685 & 0.89& -0.996473& M2: 30829 \\
\bfseries M2*& 4101.065 & 8810 & 1.07&-2.208857& M3: 30508.855 \\
\bfseries M3*& 3784.723 & 21089.53 & 2.64 &0& M4: 13200 \\
\bfseries M4& 2394.244 & Inf  &-&0& M5: 7327.616 \\
\bfseries M5& 2649.173 & Inf &- &0& FP: 27200 \\
\bfseries FP & 1987.118 & -9884.164 &- &0 &-\\
\hline
\end{tabular}
\caption{Optical specifications of the five ELT optical components in terms of: diameter, paraxial Radius of Curvature (RoC), F\#, conic constant, and relative distances between the elements. * M2 and M3 are even aspheres. Telescope specifications according to ELT ICD [\citenum{Schmid2017}]. \label{tabgeo}}
 \end{center}
 \end{table*}

\section{Telescope simulation tool}
\label{s2}

\begin{figure}
\centering\includegraphics[width=1\linewidth]{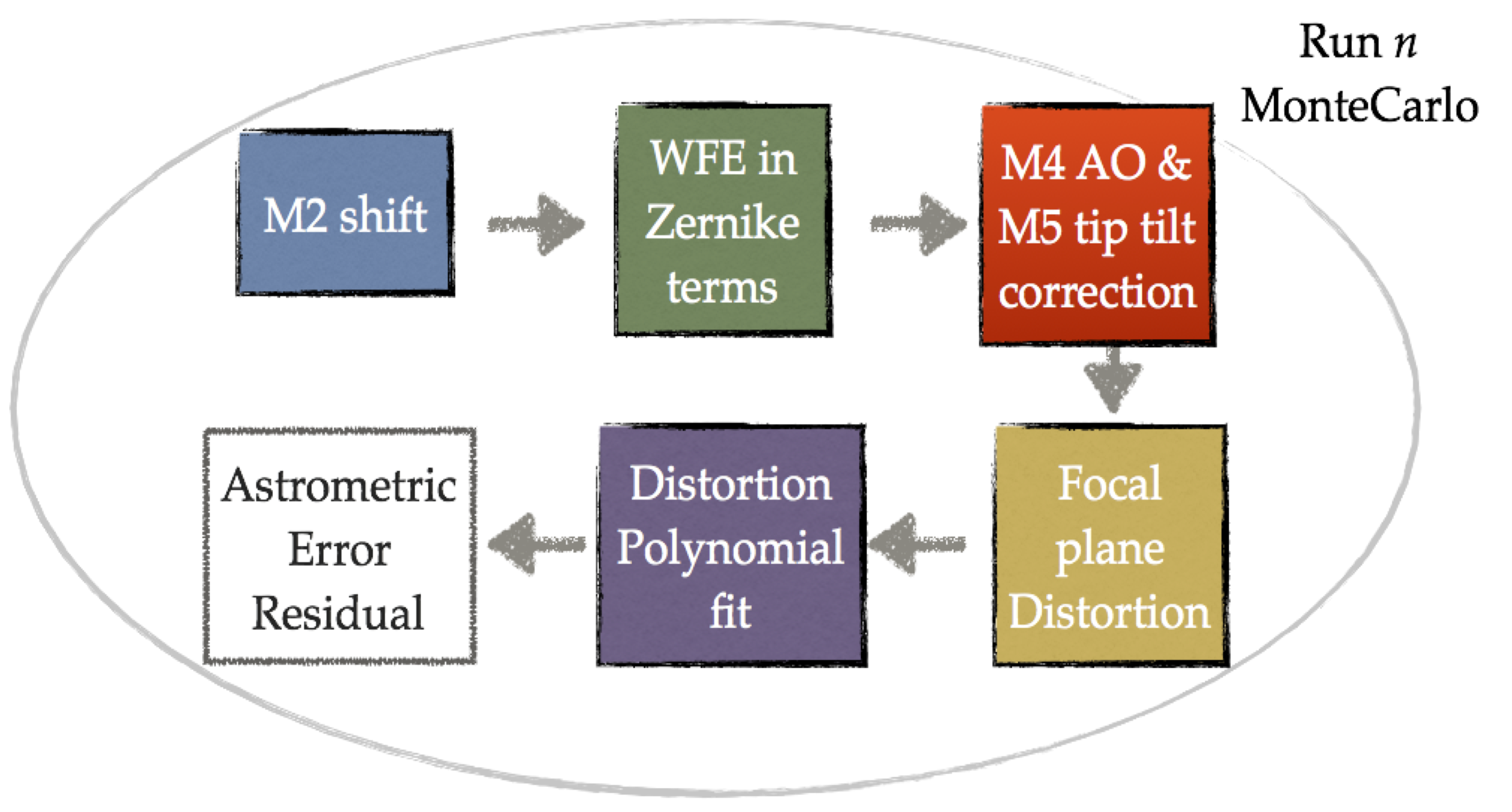}
\caption{Sequence of operations done within a single MC realization. The whole calculation (ray tracing included) is implemented in Matlab by the ZOS API interface to Zemax-OpticStudio.\label{ima2}}
\end{figure}

The approach adopted in current sensitivity study to distortions is based on Monte Carlo (MC) simulations combined with a tolerance study for the different optics of the ELT. The software used is based on the so-called ZOS API libraries that allow to control and launch simulations in Zemax-OpticStudio  [\citenum{zosapi}] from a Matlab script/environment. Figure \ref{ima2} shows the scheme of a n$^{th}$ MC simulation: (i) a certain mirror is subjected to a random positioning error within the tolerances state ($\Delta$x, $\Delta$y, $\Delta$z, $\Delta\theta$x, $\Delta\theta$y, $\Delta\theta$z) (ii) the wavefront errors (WFE) are extracted and parameterized by the first 37 Zernike terms in Noll notation [\citenum{Noll1976}] (iii) the shape of M4 is modified to minimize the residual WFE by 37 Zernike terms (iv) the telescope is refocused using the M3 active optics mechanism, and the field is steered by M5 (v) the geometric distortion at the telescope FP is extracted and fitted with a n$^{th}$ order polynomial.

The M4 correction simulated in Zemax is not a real Adaptive Optics (AO) correction because the system does not simulate any Wavefront Sensor (WFS) nor any DM influence function or correction delay; it is rather a simple minimization of the WFE to restore an acceptable SR before extracting the distortion pattern.
To accurately disentangle the effect from each optical element, for each mirror of the ELT, 20 MC random realizations are produced. The simulation has three main outputs: plate scale (PS) variation wrt. nominal design, exit pupil motion induced by the tolerances, and astrometric RMSx  \& RMSy residuals before and after the fit with 1$^{st}$, 3$^{rd}$ and 5$^{th}$ order polynomials over the whole FoV.  
The distortion pattern at the ELT FP is sampled with an equally spaced grid of 144 (12x12) points for all the MC realizations and the positions of the image points are fitted to the grid points obtained from the nominal ELT design. The latter represents the reference grid for our study and it contains a certain level of intrinsic optical distortion of the telescope nominal configuration. The grid distortion does not take into account the PSF shape and features, that consists of an additional level of complexity not taken into account for this study.
The polynomial fit expression Eq. \ref{eq2} and Eq. \ref{eq3} is the same used by \cite{Kozhurina-Platais2009}:

\begin{equation}\label{eq2}
U = A_1 + A_2X + A_3Y + A_4X^2 + A_5XY + A_6Y^2 +...+A_{21}Y^5
\end{equation}

\begin{equation}\label{eq3}
V = B_1 + B_2X + B_3Y + B_4X^2 + B_5XY + B_6Y^2 +...+B_{21}Y^5
\end{equation}

The U and V coordinates represent the grid points of the nominal intrinsic ELT distortion pattern, while the X and Y coordinates are the points from the distortion pattern of a certain MC realization. The polynomial fit is performed for 1$^{st}$, 3$^{rd}$ and 5$^{th}$ order. The 1$^{st}$ order polynomial accounts for relative translation, rotation and plate scale variations between different distortion patterns, while the 3$^{rd}$ and 5$^{th}$ order polynomials describe higher order distortions. 

\begin{table}
\begin{center}
\begin{tabular}{p{2cm}*{1}{c}} \hline
\textbf{Polynomial fit order} & \textbf{N$^\circ$ stars}\\
\hline 
\bfseries 1$^{st}$ & 3\\
\bfseries 3$^{rd}$ & 10\\
\bfseries 5$^{th}$ & 21\\
\bfseries 9$^{th}$ & 55\\
\hline
\end{tabular}
\caption{Minimum number of stars with suitable SNR for different distortion polynomial fit degrees.\label{tab_astro}}
 \end{center}
 \end{table}

\section{Distortion sensitivity analysis to telescope optics positioning errors}
\label{s3}

Official numbers for the optics positioning tolerances of the ELT are not publicly available to the scientific community yet, but ranges of tolerances for M2 and M3 are discussed by \cite{Mueller2014} and \cite{Cayrel2012} and we assume the same values also for M4 and M5 since the mirrors have comparable size. In addition, this approach gives a common metric of evaluation in a sensitivity study framework where all the mirrors are subjected to the same amplitude perturbations:

\begin{equation}\label{eq_tol}
(\Delta x, \Delta y, \Delta z) \rightarrow \pm 0.1 mm\\
(\Delta \theta x, \Delta \theta y, \Delta\theta z) \rightarrow \pm 0.01^{\circ} 
\end{equation}

\begin{figure}
\centering\includegraphics[width=1.1\linewidth]{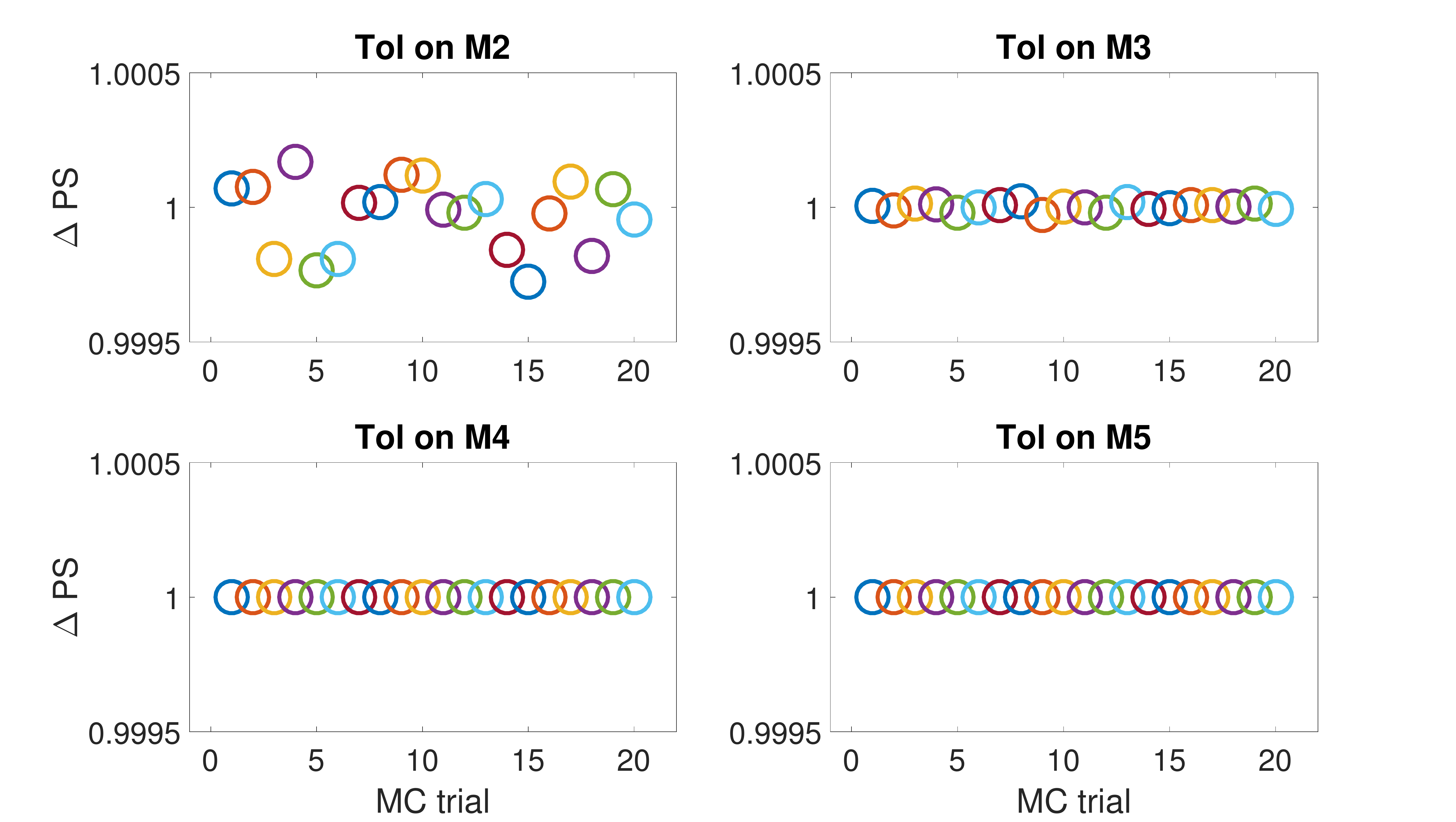}
\caption{Plate Scale relative variation wrt the nominal telescope prescription induced by the positioning errors \ref{eq_tol} for 20 MC realizations. The M2 and M3, being powered mirrors give origin to the largest perturbations. \label{ima_ps}}
\end{figure}

\begin{figure}
\centering\includegraphics[width=1.1\linewidth]{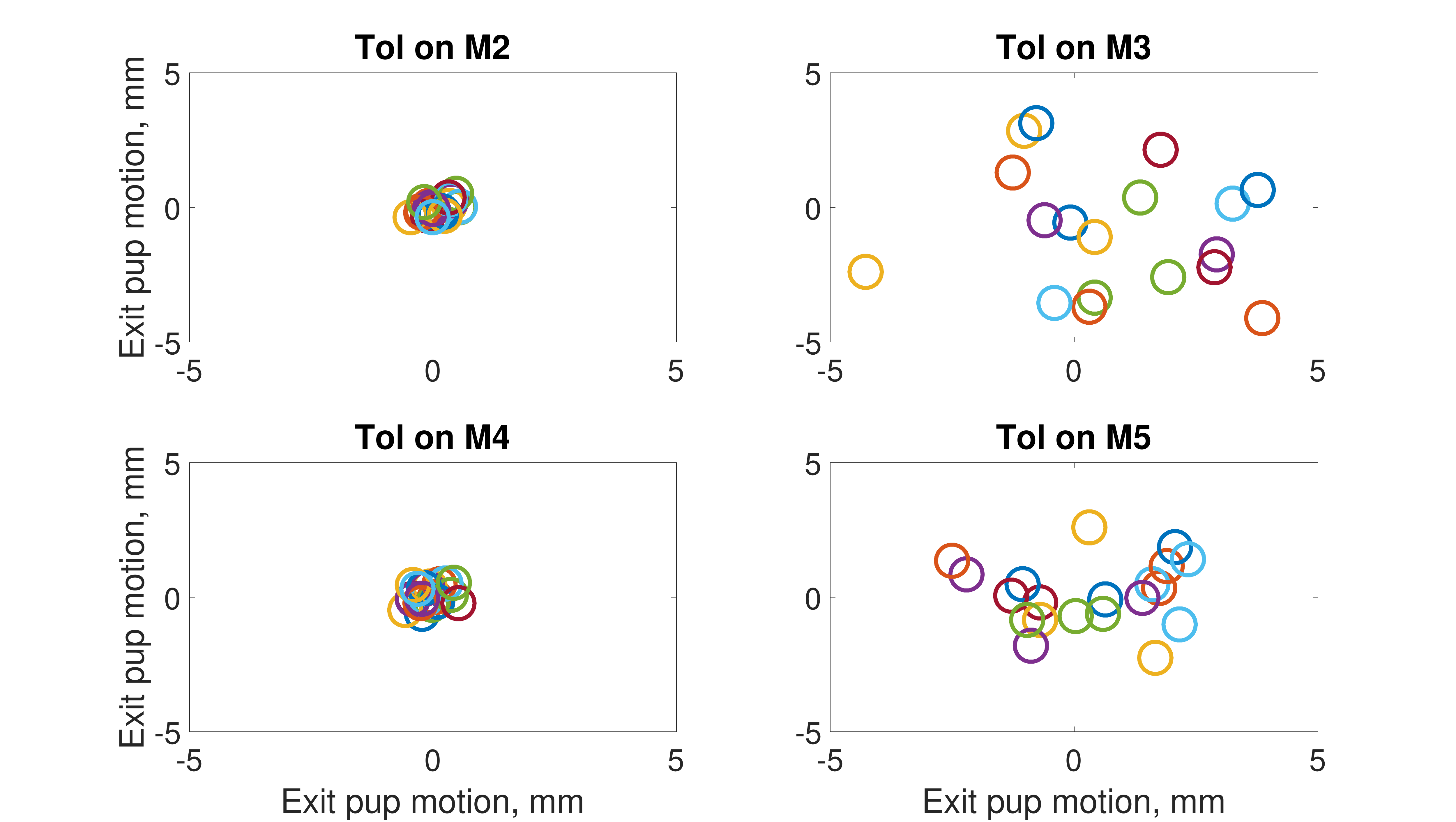}
\caption{Exit pupil motion for 20 MC realizations of the tolerance \ref{eq_tol}. The M3 and M5 creates the largest pupil displacements.\label{ima_expup}}
\end{figure}

\begin{figure}
\centering\includegraphics[width=1.1\linewidth]{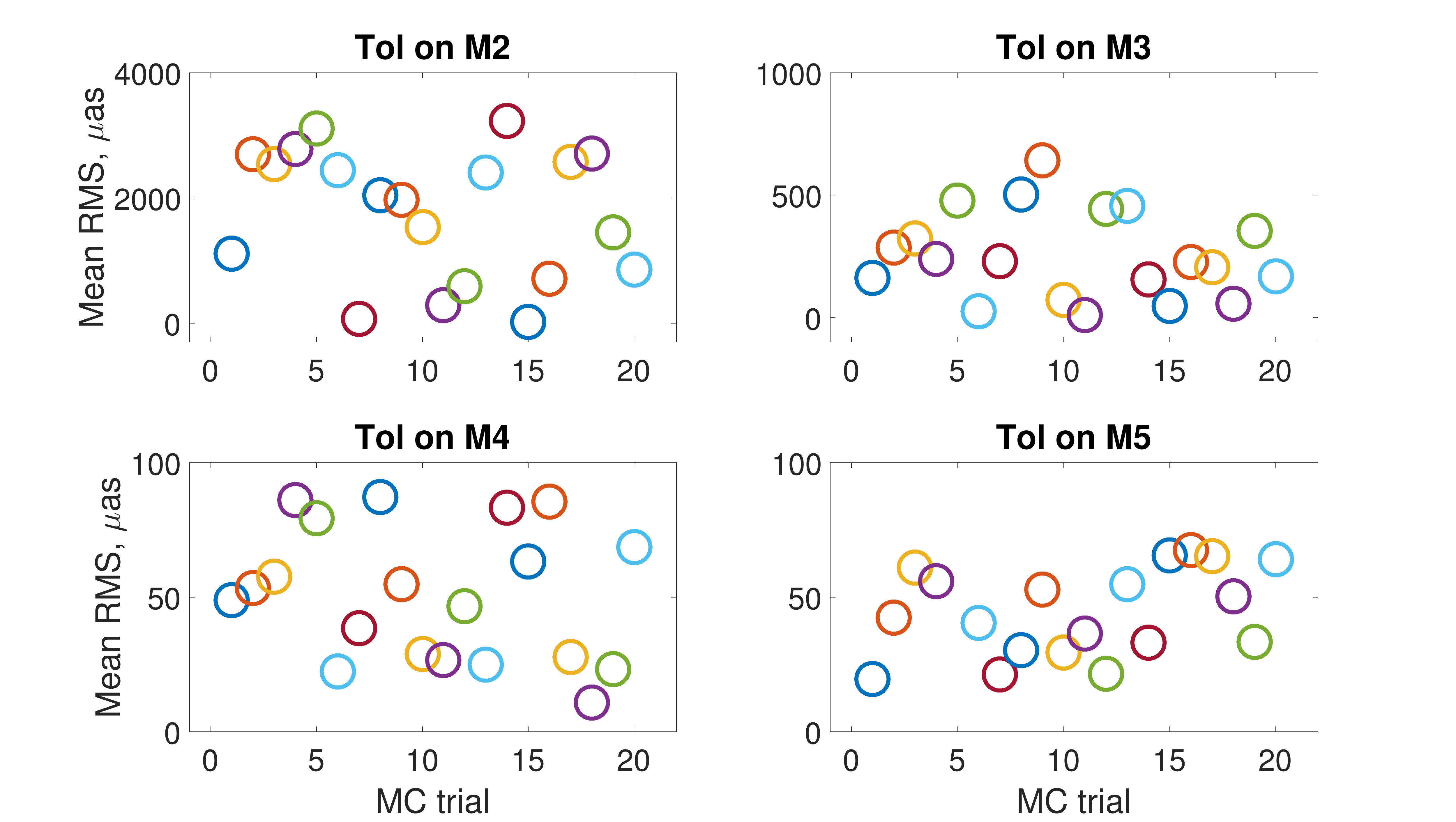}
\caption{RMS distortion over 1 arcmin FoV for 20 MC realizations before any polynomial fit. The positioning errors on the powered M2 and M3 have the largest impact on the distortion pattern. \label{ima_rmsprefit}}
\end{figure}

The $z$ axis is aligned to the optical axis between M1 and M2, $x$ is parallel to the elevation axis and $y$ completes the triad.
A total of 20 MC simulations is performed for each ELT mirror separately, each simulation picks up a random positioning error state $(\Delta x, \Delta y, \Delta z, \Delta \theta x, \Delta \theta y, \Delta\theta z)$  within the range \ref{eq_tol} and performs the operations sequence reported in the diagram of Figure \ref{ima2}. The induced PS variations, the exit pupil motion and the RMS distortion before any polynomial fit over the FoV are reported in Figures \ref{ima_ps}, \ref{ima_expup} and \ref{ima_rmsprefit} respectively.

\begin{figure}
\centering\includegraphics[width=1\linewidth]{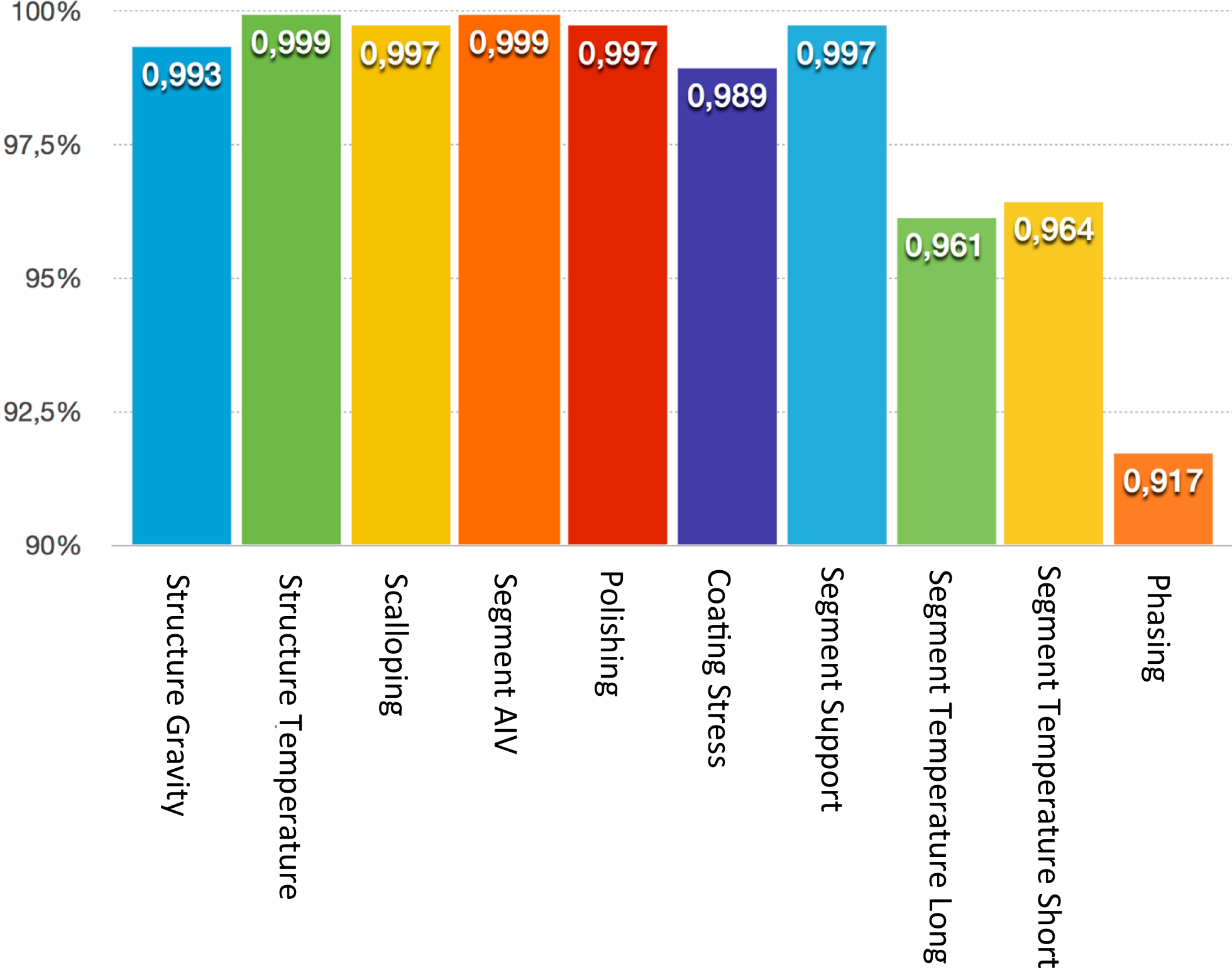}
\caption{Strehl Ratio (SR) degradation at $\lambda$ = 800 nm due to the High Spatial Frequency Errors (HSFE) originated by quasi-stationary perturbations of the ELT M1 segments. The HSFE do not introduce any differential distortions over the FoV being at the telescope entrance pupil, but they increase the astrometric error (\ref{eq1}) by decreasing the SR and broadening the PSF.\label{ima_strehl}}
\end{figure}

An exception to this analysis scheme is M1; the primary mirror of ELT being at the entrance pupil of the system does not produce field-differential aberrations and distortions over the FoV. The astrometric systematics induced by M1 relate mainly to the decrease of the SR caused by the high spatial frequency errors (HSFE) of the M1 segments as shown in the histogram of Figure \ref{ima_strehl}. Among ten different HSFEs, the phasing errors of the M1 segments produce the largest aberrations resulting in a RMS WFE $\sim$ 37 nm \citep{Marchetti2015}.


\begin{figure}
\centering\includegraphics[width=1.1\linewidth]{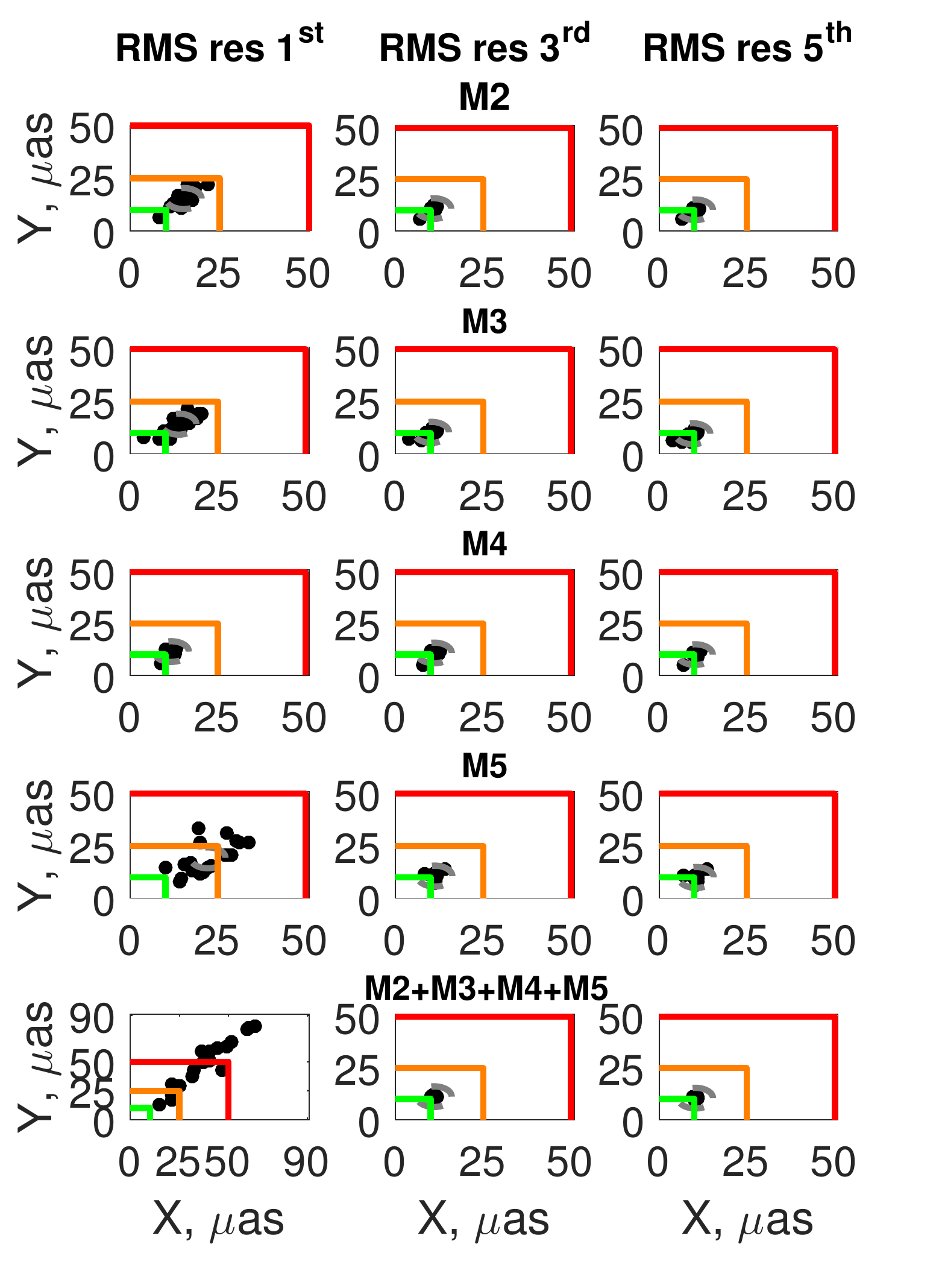}
\caption{RMS distortion over 1 arcmin FoV for 20 MC realizations after 1$^{st}$, 3$^{rd}$ and 5$^{th}$ order polynomial fit. Comparing with Figure \ref{ima_rmsprefit}, the reader can note how the ELT distortions are dominated by PS changes being the 1$^{st}$ order RMS residual fit already significantly smaller than 50 $\mu as$. The dispersion of the different MC is about 5 $\mu as$ (grey circle). Combining all the tolerances together  (Bottom) some MC show distortion residuals $>$ 50 $\mu as$ post 1$^{st}$ order fit. The simulations indicate also that the ELT distortion extends up to 3$^{rd}$, but not at 5$^{th}$, having the RMS residuals post fit the same amplitude in both cases. \label{ima_rms_postfit}}
\end{figure}

The other ELT mirrors, M2, M3, M4, M5 cause distortion at different levels; the plots in Figure \ref{ima_rmsprefit} shows the RMS residuals distortion of the MC simulations before the polynomial fit is applied. The worst offender in terms of optical distortions is the M2, and among the different positioning errors, the axial shift wrt M1, $\Delta z$, is the most problematic perturbation that causes fast and large variations of the PS. Although more than one scenario is possible, we assume that M3 is compensating for the defocus induced by the axial shift of M2. The M3 in fact is equipped with an active optic mechanism that allows to change the focus of the telescope \cite{Cayrel2012}. A $\Delta z(M2) = 0.1$ mm leads to a PS variations $\sim$ 0.02\% after refocus with M3, that translates into a $\sim$ 4-5 mas drift of the field distortions over an arcmin FoV. This estimate is in good agreement with the ELT Interface Control Document (ICD) specifications [\citenum{Schmid2017}]. The field distortions induced by the M3 are about an order of magnitude smaller than those of the M2 and one order of magnitude greater than those of the M4 and M5. The reason of this difference is due to the fact that M2 and M3 are powered mirrors with fast f/\# while M4 and M5 are flat. In this work the shape residual errors and the mid-spatial frequency errors (MSFEs) of the mirrors are not considered. The magnitude of the MSFEs depends on the manufacturing process of the optics and the size of the tool used for grinding-polishing the surfaces. Ultra precise optics for three mirror anastigmat systems can achieve residual MSFEs in the order of 20 nm PV \citep{Scheiding2010}. Another factor to be considered is the position of the optics with respect to the focal plane: surfaces close to the focal plane where the light beams from different field points are converging and their footprints are significantly apart can originate high order distortions that require polynomials of order $\geq$ 7$^{th}$ to be fit \citep{Rodeghiero2018}. In the ELT case, there are no optics close to the focal plane where the MSFEs can play an important role. Once the polynomial fit (Eq. \ref{eq2} and \ref{eq3}) is applied to the distortion patterns from each MC simulation the residual RMS distortion over 1 arcmin FoV can be assessed as reported in Figure \ref{ima_rms_postfit}. The RMS residual distortion over the FoV is shown for each MC realization in comparison to the typical post-processing astrometric precision requirement ($\sigma \sim$50 $\mu$as) of instruments like MICADO \cite{Davies2016} and IRIS \cite{Larkin2016}. As the reader can observe, already the 1$^{st}$ order fit leads to small astrometric residuals mostly below the instruments requirement, meaning that predominant ELT distortions are caused by plate scale variations, translations and rotations. The combination of the positioning errors on all the telescope optics (Fig. \ref{ima_rms_postfit}, bottom) gives origin to a greater dispersion of the MC realizations (post 1$^{st}$ order fit) with the $\sim$ 50\% of them falling beyond the threshold of 50 $\mu$as.  
The 3$^{rd}$ order fit further decreases the RMS residuals leading to a compact cloud of RMS points with a centroid around 12-13 $\mu$as for all the mirrors, while higher order polynomials like e.g. 5$^{th}$ order does not bring any additional improvement. This result is of crucial importance: the ELT distortion pattern is dominated by low order distortion modes and since a 3$^{rd}$ fit already breaks down the RMS residuals to 12-13 $\mu$as, only a relatively small numbers of stars $\sim$ 10, is required to perform the polynomial fit of the distortion pattern (Table \ref{tab_astro}).

\begin{figure}
\centering\includegraphics[width=1.1\linewidth]{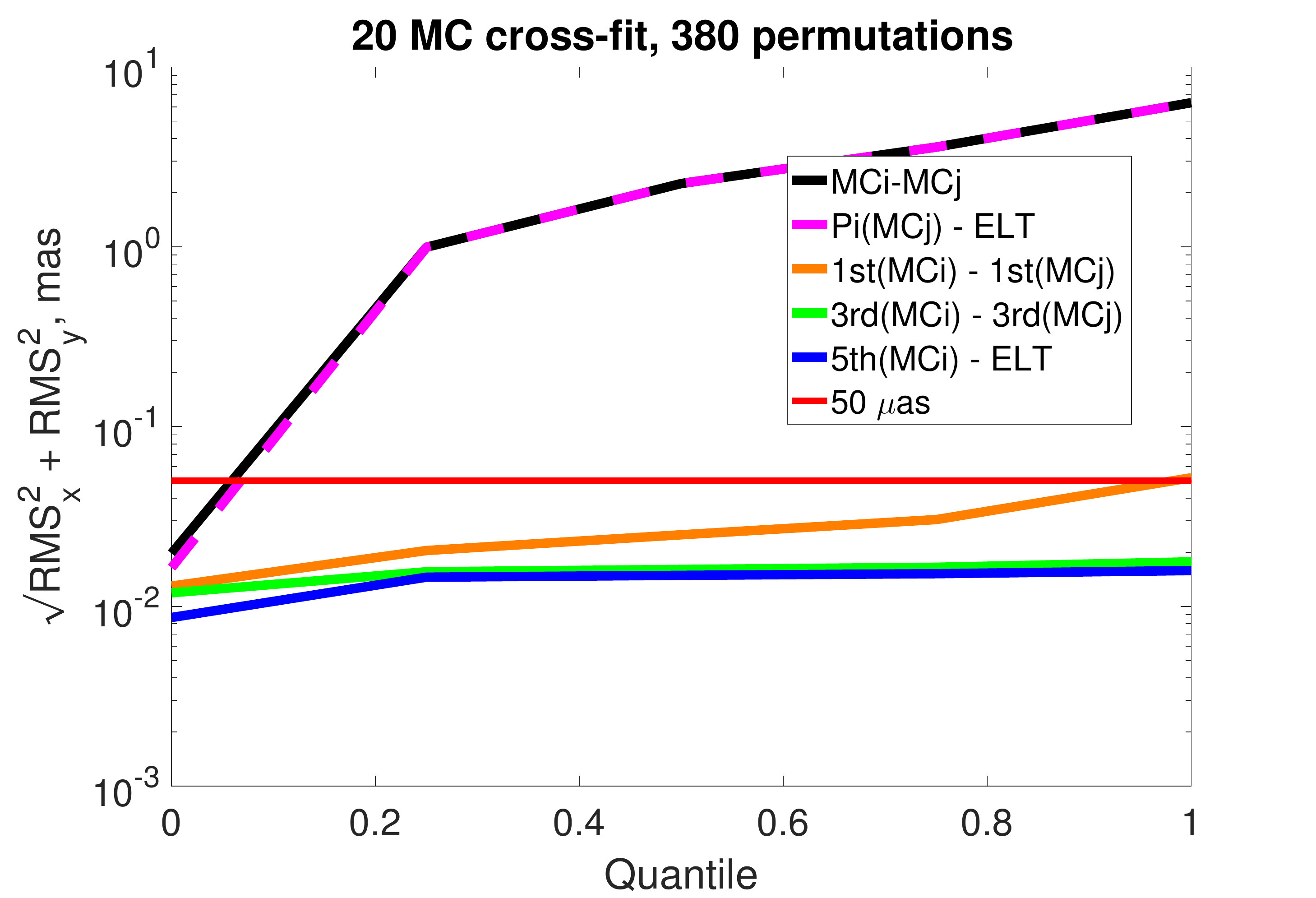}
\caption{Cross-check of the RMS residual distortions of 20 MC realizations fitted with sets of interchanged polynomial coefficients (magenta). The high residual between the fit of a certain MC realization and the nominal ELT distortion, is due to the PS mismatch, that once removed with a 1$^{st}$, 3$^{rd}$ or 5$^{th}$ fit (orange, green, blue), returns residuals below the 50 $\mu$as. A certain distortion solution of the telescope can hardly be applied to another epoch, suggesting the need for frequent on-sky calibrations of the PS.\label{ima_crossfit}}
\end{figure}

An important additional aspect of the tolerance study is the time domain over which these opto-mechanical instabilities take place. Positioning errors in the range of tolerances of Eq. \ref{eq_tol} are originated mainly by gravity flexures, thermal gradients and wind disturbances. The thermal gradient follows the nighttime timescale profile $\sim$ hour, the gravity flexures and torques on the telescope mirrors follow the typical timescales of the pointing and tracking, $\sim$ minutes, while the wind disturbances are faster, typically $\sim$ seconds. Although the WFE aberrations coming from these perturbations are efficiently corrected by the AO system, the physical displacement of the mirror creates also a certain level of optical distortion that challenges the astrometric precision. 
The rate of change of such disturbances determines how frequently the observer needs to calibrate the telescope on sky. In Figure \ref{ima_crossfit} we report the average post-fit RMS$_{x,y}$ residual distortion interchanging the polynomial coefficients (eq. \ref{eq2}) from different MC realizations in a random series of permutations to assess how well the polynomial coefficients $P_i$ obtained with the fit of the distortion pattern from the $MC_i$ simulation can fit the distortions of another realization $MC_j$, i.e. $P_i(MC_j)$ with i $\neq$ j.

The polynomial coefficients interchange returns RMS residuals between 0.1-6 mas (magenta line) at the same level of the direct subtraction of the distortion patterns without any polynomial fit (black line), highlighting the fail of the fit in reproducing the ELT nominal distortion pattern. This large residual is due to the large PS mismatch between two random MC realizations that is the predominant term of the distortion pattern. In fact, the RMS residual of post 1$^{st}$, 3$^{rd}$ and 5$^{th}$ fit (orange, green, blue) between random MC permutations is smaller than 50 $\mu$as for all the permutations. Figure \ref{ima_crossfit} indicates that: (i) a certain distortion solution of the telescope can hardly be applied to another epoch given the high rate of change of PS, suggesting the need for frequent on-sky calibrations of the low order distortion terms; (ii) the variations of the distortions of higher order with respect to PS is very small and barely sensitive to the tolerances \ref{eq_tol}.  
In the next two sections \ref{s4} and \ref{s5} we discuss two ELT typical operation scenarios where the positioning errors caused by the dynamical effects of gravity and wind disturbances induce changes of the distortion pattern at seconds and minutes timescales.

\section{Telescope Low Order Optimization loop}
\label{s4}

\begin{figure}
\centering\includegraphics[width=1.1\linewidth]{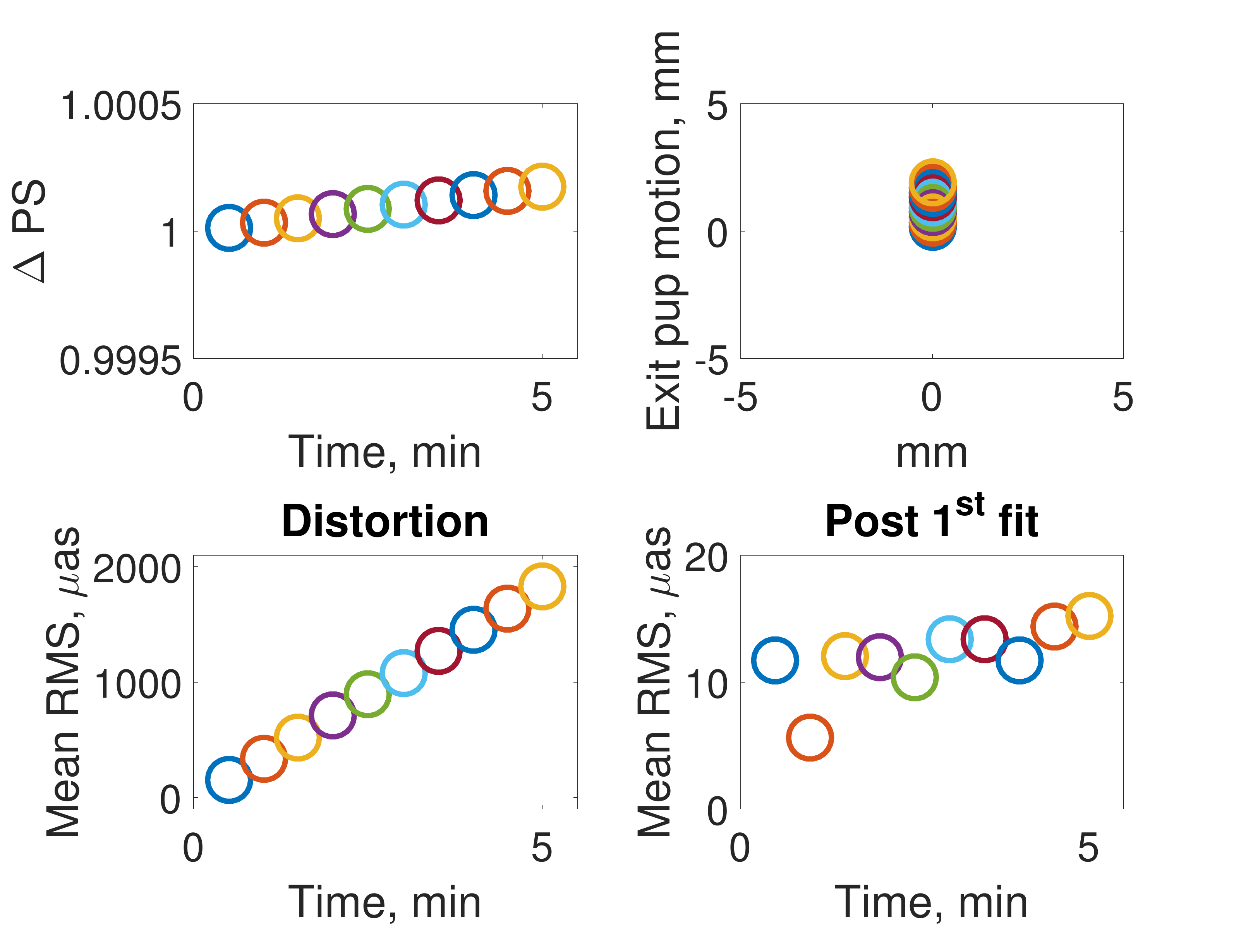}
\caption{PS variation, exit pupil motion and RMS distortion during a typical interval of 5 minutes between two consecutive LOO loops. Assuming a 120 sec exposure the RMS distortion can rise up to $\sim$ 1 mas. The distortion pattern is dominated by the plate scale term linearly increasing with the axial shift of M2 and efficiently removed already by the 1$^{st}$ order fit. \label{loo_dist}}
\end{figure}

The MC tolerance study carried out in Section \ref{s3} points out that M2 is the most sensitive element to opto-mechanical misalignments in the ELT telescope thus producing significant distortions over 1 arcmin FoV. \cite{Mueller2014} reports some scaling relations of interest for the current study. The gravity flexures between the M1 and M2 change at a rate of $dy,dz \sim 0.5$ mm/h and the axial positioning error produces a defocus RMS WFE that scales as $dZ^0_2 \sim$ 0.39 $\mu$m/mm. The thermal expansion of the telescope structure varies as $dz \sim$ 0.36 mm/K. Combining in quadrature the drifts we find that already in 5 minutes the axial drift accumulates to $dz \sim$ 50 $\mu$m, and this has an impact in terms of distortions comparable to the tolerances \ref{eq_tol}. The current control strategy of M2 according to \cite{Mueller2014} foresees a passive supporting structure with wiffletrees without an active optic mechanism, rather the structure is built to have the maximum errors repeatability under gravity torque perturbations. The M2 is left drifting for a period of 5 minutes while the WFE is compensated by the AO system; afterwards, the M1-M2 collimation is restored by the so called Low Order Optimization loop (LOO) that repositions the M2 back to its optimal position. The LOO loop is actuated both to avoid the saturation of the DM dynamic range and to avoid an uncontrolled growth of aberrations and distortions. This dynamical effect constitutes a challenge to the astrometric observations being the M2 physically drifting with respect to the M1 and inducing a distortion variation over the field. Figure \ref{loo_dist} shows the result of intra-LOO distortion drift: a series of multiple, progressively increasing, scollimation steps of M2 are simulated over a time frame of 5 min for a telescope zenith angle of 45$^\circ$. 

At each step the positional and angular drift of M2 is linearly increased along the optical axis $z$ an the gravity vector $y$ to reach $\Delta(z,y)_{max} \sim$ 0.07 mm $\rightarrow  \sqrt{\Delta y^2 + \Delta z^2} \sim$ 0.1 mm and $\Delta \theta_x\sim$ 0.01$^{\circ}$ at the end of the 5 min interval. During the simulated M2 drift the WFE is compensated with the ELT-M4 DM using the first 37 Zernike polynomials. The RMS distortion drift is driven by the PS drift induced by the M2 axial motion and it can reach $\sim$ 2 mas/arcmin in 5 minutes. During typical NIR image exposure times with ELT, $\sim$ 120 sec, the RMS distortion is expected to be $\sim$ 1 mas, largely due to PS. While the refocus of the telescope is guaranteed by the combined action of M3 and M4, to contain the PS drift during the exposure another DM is required, meaning $de$ $facto$ that $\mu$as astrometry is enabled only with a Multi Conjugated AO system (MCAO). In the next paragraph we discuss a specific case of study where an MCAO system is combined with the ELT to enhance the astrometric performances over a wide FoV.

\section{Wind perturbation on M2}
\label{s5}

\begin{figure}
\centering\includegraphics[width=1.1\linewidth]{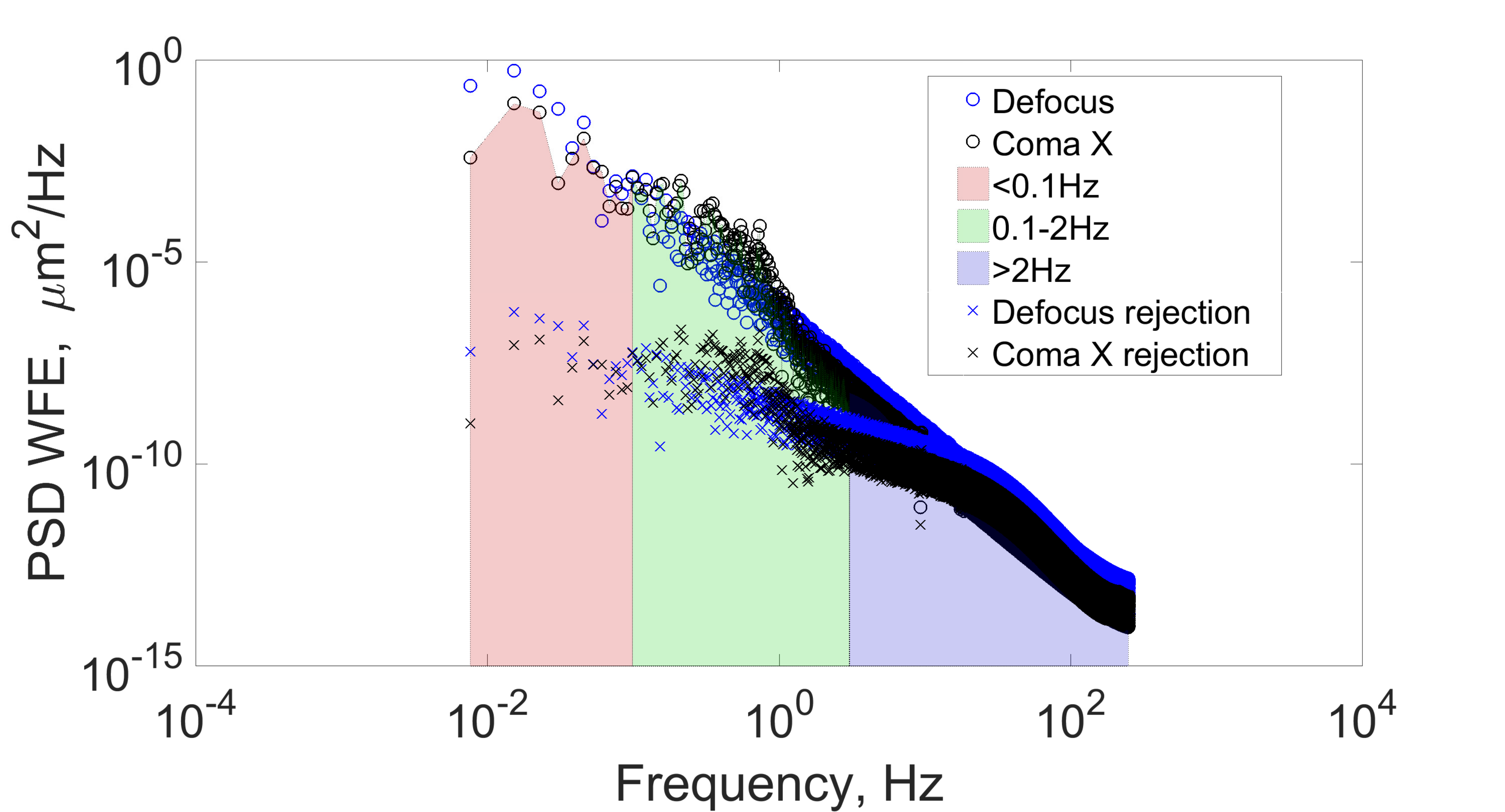}
\caption{PSD of the expected defocus and coma-X WFE (empty circles) induced by the wind on the ELT-M2 \citep{Marchetti2015}. Residual defocus and coma-X PSD after having applied a typical AO rejection function (Eq. \ref{m4_rejf}) for ELT-M4.\label{wind_on_m2}}
\end{figure}

The effect of the wind load on the telescope structure has been evaluated with Ansys CFD simulations~\cite{Tamai2014}. The wind speed around the M2 is expected to be $\sim$ 8-12 m/s (\cite{Tamai2014}, \cite{Marchetti2015}). The wind perturbation is expected to influence also the shape of the mirror itself (not considered in this paper): the rms mirror deformation is approximately proportional to the square of the wind velocity~\citep{Cho2001}.
To assess the distortion induced by the M2 motion under the wind perturbation, we produce a series of MC simulations with different M2 position and tilt offsets to reproduce a comparable WFE at the exit pupil as reported in the ESO dataset \cite{Marchetti2015}. 
The M2 positioning errors that reproduce such WFE are $\sigma \sim 0.25\times(\Delta z)$ those assumed in Section \ref{s3} (Eq. \ref{eq_tol}). The AO correction compensates very efficiently the WFE as shown in Figure \ref{wind_on_m2}, here e.g. the low order terms defocus and coma X, already with a single deformable mirror (ELT-M4). 
%

 \begin{figure}
\centering\includegraphics[width=1.\linewidth]{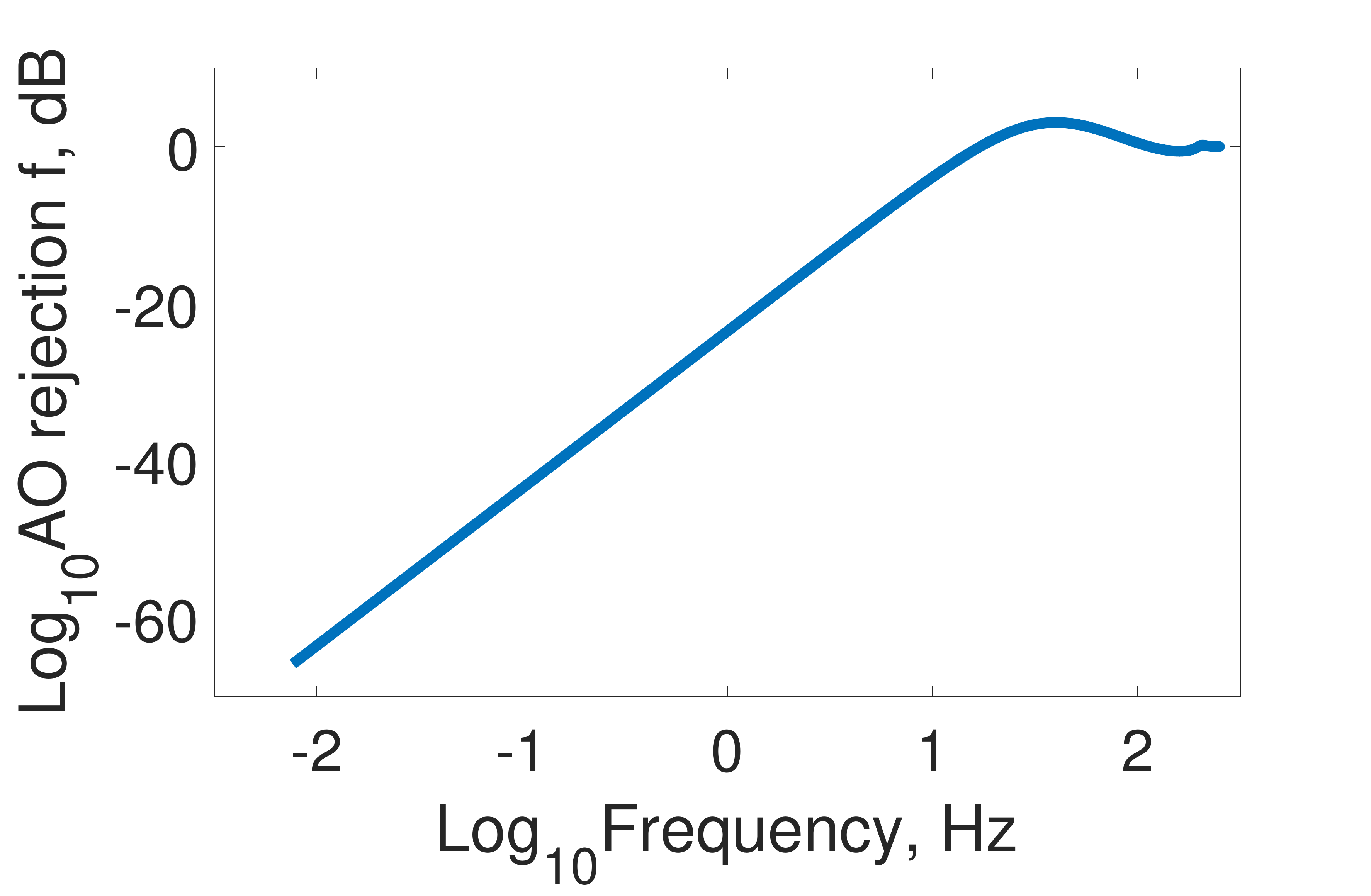}
\caption{AO rejection function for the ELT-M4 (Eq.\ref{m4_rejf}). \label{AO_rej}}
\end{figure}

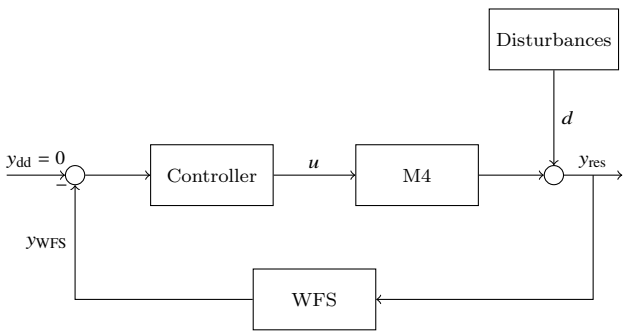
\begin{figure}
\tikzstyle{block} = [draw, fill=white!20, rectangle, 
    minimum height=3em, minimum width=6em]
\tikzstyle{sum} = [draw, fill=white!20, circle, node distance=1cm]
\tikzstyle{input} = [coordinate]
\tikzstyle{output} = [coordinate]
\tikzstyle{pinstyle} = [pin edge={to-,thin,black}]

\hspace*{-1.5em}\begin{tikzpicture}[every node/.style={scale=0.9},auto, node distance=2cm]
    \node [input, name=input] {};
    \node [sum, right of=input] (sum) {};
    \node [block, right of=sum] (controller) {Controller};
    \node [block, right of=controller, node distance=3cm] (system) {M4};
    \node [sum, right of=system, node distance =2cm] (sum1) {};
    \node [output, right of=sum1,node distance =1cm] (output) {};
	\draw [->] (controller) -- node[name=u] {$u$} (system);
    \node [block, below of=u] (measurements) {WFS};
    \node [block, above of=sum1] (disturbances) {Disturbances};

    \draw [draw,->] (input) -- node {$y_{\mathrm{dd}}=0$} (sum);
    \draw [->] (sum) -- (controller);
    \draw [->] (system) -- (sum1);
    \draw [->] (sum1) -- node [name=y] {$y_{\mathrm{res}}$}(output);
    \draw [->] (y) |- (measurements);
    \draw [->] (measurements) -| node[pos=0.99] {$-$} node [near end] {$y_{\mathrm{WFS}}$} (sum);
    \draw [->] (disturbances) -- node[name=u] {$d$}(sum1);
\end{tikzpicture}
\caption{Control structure of the ELT AO system for the Zernike modes defocus and coma X. }
\label{plot:controlstructure}
\end{figure}

The AO-M4 rejection performance (Figure \ref{AO_rej}) can be determined considering the dynamic behavior of the AO system, which is represented in  Fig.~\ref{plot:controlstructure} schematically. The loop speed of the AO system is set to 500 Hz.

The residual wavefront error $y_{\mathrm{res}}$ between disturbances $d$ and the M4 wavefront $y_{\mathrm{M4}}$ is measured by the wavefront sensor $y_{\mathrm{WFS}}$. Due to exposure and reconstruction of the wavefront the measurements are delayed by one or two sample depending on the sample rate. In the discrete frequency domain the transfer function $G_{\mathrm{WFS}}$ is described by 
\begin{align}
	G_{\mathrm{WFS}}(z) = z^{-2},
\end{align}
where $z$ corresponds to $e^{j \omega T_\mathrm{s}}$ in the steady state, $\omega$ is the circular frequency and $T_s$ the sample time of the AO loop. The mirror dynamic in the defocus and coma X mode is described by a second order system
\begin{align}
	G_{\mathrm{M4}}(z) = \frac{0.7552 z^2 + 1.51 z + 0.7552}{ z^2 + 1.357 z + 0.6635}.
\end{align}

The model parameters are adjusted to the the tip-tilt transfer function from ~\cite{Sedghi2010}, because we do not have the current transfer function of defocus and coma X. This assumption is allowed for the considered frequency range. The control goal is to obtain a desired wavefront $y_{\mathrm{dd}}$ without disturbances, which is set to zero for simplification. The wavefront is controlled in Zernike modes, where we use an integral controller for both modes without considering coupling effects. Due to the used mirror dynamics we also assume the controller design from ~\cite{Sedghi2010}
%
%

\begin{align}
	G_{\mathrm{C}}(z) = 0.094 \frac{z+1}{z-1}.
\end{align}

Based on the introduced models we can derive the M4 rejection function for the WF rms (Fig. \ref{AO_rej})

\begin{align}\label{m4_rejf}
G_{\mathrm{RF}}(z) = \frac{Y_\mathrm{res}(z)}{D(z)} = \frac{1}{1+G_{\mathrm{WFS}}(z) G_{\mathrm{C}}(z) G_{\mathrm{M4}}(z)}.
\end{align}

The residual PSD $S_{\mathrm{y_{res}}}$ (crosses in Fig. \ref{wind_on_m2}) can be calculated by 

\begin{align}
S_{\mathrm{y_{res}}}(\omega) = |G_{\mathrm{RF}}(e^{j \omega T_\mathrm{s}})|^2 S_{\mathrm{d}}(\omega),
\end{align}

where $S_{\mathrm{d}}(\omega)$ is the input PSD of defocus or coma X in Fig.~\ref{wind_on_m2} (empty circles). 

\begin{figure}
\centering\includegraphics[width=1.05\linewidth]{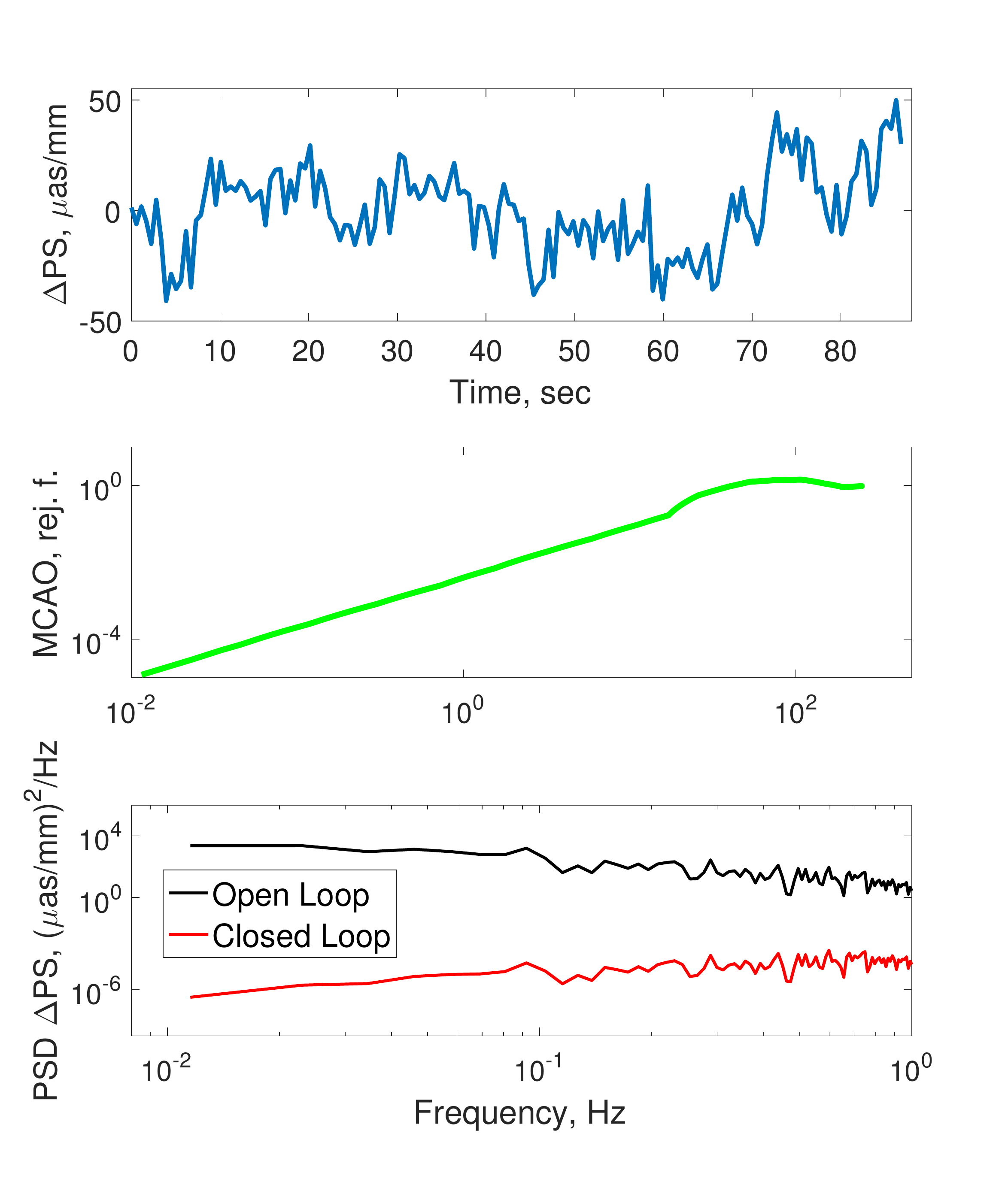}
\caption{\textbf{Top} Simulation of the expected PS variation for wind perturbation on M2. \textbf{Middle} Rejection function derived from the MCAO simulator. \textbf{Bottom} Open and closed loop PSD of the $\Delta$PS in the MCAO simulator.\label{ima_mcao_rej}}
\end{figure}

Although the AO correction compensates very efficiently the WFE, the mirror offset causes a distortion variation that challenges the astrometric observations. Being the wind disturbances a dynamic effect subjected to fast changes, also the correction of the distortions (mostly low oder) has to come from the AO system. While a single deformable mirror cannot control both WFE and plate scale, a multi-conjugated AO (MCAO) system can stabilize also the latter systematic. The MCAO system is modeled based on the current ELT scheme that foresees a high order adaptive mirror (M4) conjugated at 625 m above the entrance pupil. A second deformable mirror conjugated at 15 km is added to create an MCAO system and both deformable mirrors are described in terms of a Zernike polynomial modal base. Nine natural guide stars are uniformly distributed over a FoV of 2 arcmin and for each star a wavefront sensor. To control the PS variations 3 stars are sufficient, but to ensure a good seeing correction all over the FoV other 6 stars are added to the AO loop. This configuration is in the ball park of the envisaged first-generation AO systems for the ELTs.  We run an End-2-End simulation of the system at a AO loop frequency of 500 Hz on a seeing of 0.65 arcsec.
%
%
On top of the atmospheric wavefront perturbation we add a PS modulation estimated from the wind disturbances on M2 (Top Fig. \ref{ima_mcao_rej}).
Given the purpose of our analysis, which focuses rather on the dynamical response of the AO control to the disturbance than to the noise, neglecting the effect of wavefront sensor noise that is modeled as a simple linear first derivative sensor. On the corrected wavefront we finally compute the residual plate scale and we build the open loop and closed loop Power Spectrum Density (PSD) of the PS (Bottom Fig.\ref{ima_mcao_rej}). The MCAO rejection function is computed from the open and closed loop PSD as shown in Figure~\ref{ima_mcao_rej} (Middle).\\

The average Strehl in $K$-band over the full FoV is $\sim$ 5-10 \% obtained with the first 37 Zernike modes of the DMs. The simulation indicates that the MCAO correction is able to fully stabilize the plate scale due to the axial positioning error of M2 originated from the wind perturbation. The open loop PS rms is $\sigma_{OL}(\Delta PS) \sim 15$ $\mu$as/mm in the range of 0.02-2 Hz (Top Fig. \ref{ima_mcao_rej}), while residual PS in closed loop is reduced by three orders of magnitude to  $\sigma_{CL}(\Delta PS) \sim 0.15$ $\mu$as/mm. These numbers translated in PSF jitter at the edge of 1 arcmin FoV are $\sigma_{OL} \sim 5$ mas/arcmin and $\sigma_{CL} \sim 0.041593$ mas/arcmin.
This residual is of course well below the upper limit due to the pure temporal delay we may set for a PS disturbance with a typical frequency of a few Hz and a MCAO loop correction running at 500 Hz. Therefore, we can conclude that the temporal power spectral density of the M2 distortion disturbance due to wind shake should be rejected by an MCAO control at the levels of tip-tilt residual rms. If that is assumed to be 1/10$^{th}$ of the FWHM, in $H$-band we can expect $8.5\cdot \sqrt{2}/10$ mas over a typical MCAO NGS distance of 2 arcmin, which translates into a PS stability of 0.6 mas/arcmin for good AO performance, and would bring the telescope-induced PS jitter just to the level significantly smaller than the diffraction limited PSF, needed to not compromise the astrometric performance (Eq.~\ref{eq1}). In single-conjugate AO mode however, the Strehl and astrometric performance at the edge of an arcmin sized field would be seriously compromised by the telescope-induced plate-scale jitter, which would probably prevent from taking advantage of having an M1 aperture comparable to the outer scale of the atmosphere (\cite{Clenet2015} estimated that even in SCAO the ELT NIR SCAO corrected PSF could have non-negligible Strehl).

\section{Telescope induced field rotation}
\label{s6}

The combination of two or more plane mirrors with relative tilt changes the image orientation of the object seen through the system \citep{Baker1928}. This effect is normally exploited in the so-called k-mirror device to derotate sky images against the sidereal motion at Alt-Az telescopes focus \citep{Guo2014}. The ELT M4-M5 duo is expected to induce field rotation by two different mechanisms: uncontrolled opto-mechanical tolerances (\ref{eq_tol}) of the mirror cells and tip-tilt AO correction of the M5 that compensates the atmospheric image jitter. The reflection matrix notation is particularly efficient in calculating the image orientation from an ensemble of plane mirrors as a simple matrix product applied to the ray coordinates vector $k$ in a given reference frame:    

\begin{equation}\label{eq_mirrormat}
k_2 = M_3M_2M_1k_1 = M_{eff}k_1
\end{equation}

In this formalism, each mirror is represented by a matrix $M_i$ (Eq. \ref{eq_matrices}) that accounts for its orientation state in the reference frame. The order of multiplication between the matrices and the ray vector follows the same order of the ray path within the system. The explicit form of the reflection matrix is given by \cite{Walles1964}:

\begin{equation}\label{eq_matrices}
\left(
\begin{array}{c}
l\\
m\\
n\\
\end{array}
\right)
= 
\left(
\begin{array}{ccc}
1 - 2L^2,\hspace{0.2cm} -2LM,\hspace{0.2cm} -2LN\\
-2LM,\hspace{0.2cm} 1 - 2M^2, \hspace{0.2cm}-2MN\\
-2LN,\hspace{0.2cm} - 2MN, \hspace{0.2cm}1-2N^2\\
\end{array}
\right)
\left(
\begin{array}{c}
l_0\\
m_0\\
n_0\\
\end{array}
\right)
\end{equation}

where $(l,m,n)$ and $(l_0,m_0,n_0)$ are direction cosines of the reflected and incident ray respectively, and $(L,M,N)$ represent the direction cosines of the normal to the mirror surface.

\begin{figure}
\centering\includegraphics[width=0.7\linewidth]{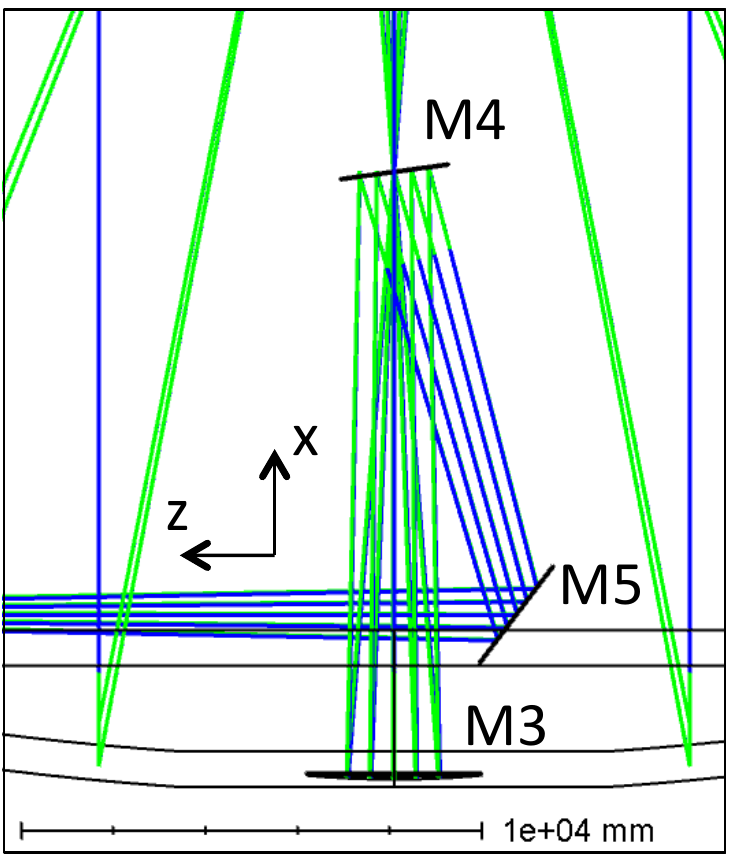}
\caption{Zoom view of the ELT M3, M4 and M5 unit: M4 is tilted by $\theta_y = 7.75^\circ$ and M5 by $\theta_y = 37.25^\circ$. The $y$ axis is perpendicular to the page plane. \label{ima_kmirror}}
\end{figure}

\begin{figure}
\centering\includegraphics[width=1\linewidth]{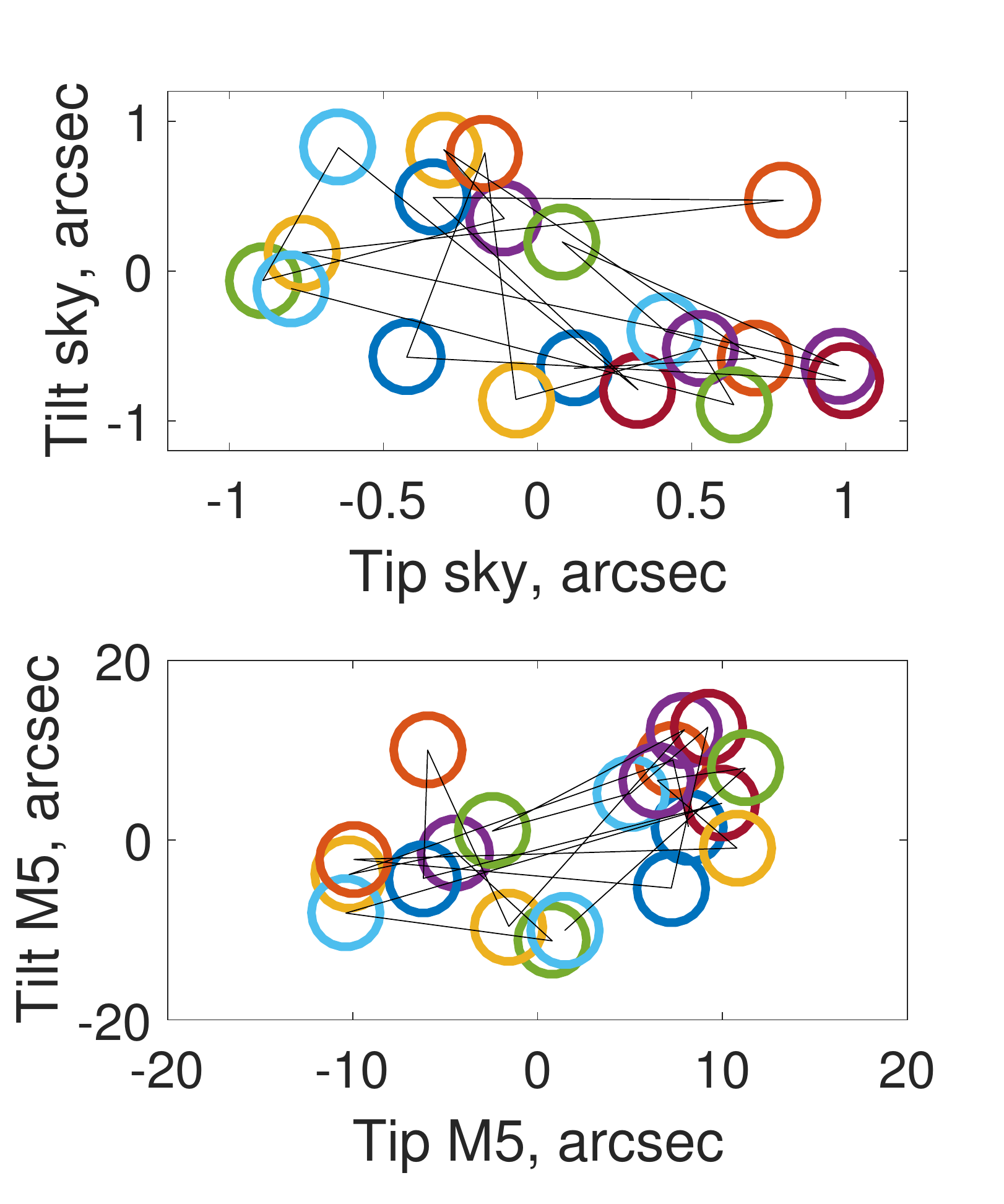}
\caption{\textbf{Top:} simulation sequence of 20 seeing tip-tilt jitter configurations used to estimate the field rotation induced by ELT-M5 while compensating the atmospheric seeing jitter. \textbf{Bottom:} corresponding amplitude of tip-tilt correction required from M5 to compensate the above seeing jitter. \label{ima_tip_tilt}}
\end{figure}

The impact on the Strehl due to residual field rotation (FR) increases progressively with the width of the FoV and it poses important challenges to the astrometric observations by smearing out the PSF in the outer part of the field.  Figure \ref{ima_kmirror} shows the layout of the ELT-M3, M4 and M5 unit: M4 and M5 have a tilt in the plane of the image of respectively $\theta_y = 7.75^\circ$ and $\theta_y = 37.25^\circ$. Tilts around $y$ axis to do not create FR, while the rotations around $x$ and $z$ induce FR as shown in Table \ref{tab_FR} for typical opto-mechanical tolerances in the range \ref{eq_tol}. The AO M5 correction is used to compensate the atmospheric PSF jitter with a rate up to 100 Hz, but this correction causes also FR and smearing of the off-axis PSFs. Assuming a random $\theta_{seeing} = 1"$ seeing tip-tilt jitter distribution as shown in Figure \ref{ima_tip_tilt} (Top), we calculate the tip-tilt correction amplitude that needs to be applied to M5 for stabilizing the field by means of Equation \ref{eq_m5}:

\begin{equation}\label{eq_m5}
 \theta_{M5} = \frac{1}{2}\arctan{\frac{\theta_{seeing}\times PS}{BFD}}
\end{equation}

With BFD the back focal distance of the ELT. The M5 tip-tilt correction corresponding to the seeing perturbation is shown in Figure \ref{ima_tip_tilt} (Bottom). The FR amplitude for the correction of $\sim$ 1" seeing estimated with the reflection matrices (Eq. \ref{eq_mirrormat}) is in the order of $\sim$12" (Table \ref{tab_FR}) thus producing a PSF jitter at edge of 1 arcmin FoV of $\sim$ 2.4 mas. This number is cross-checked independently with a non-sequential Zemax-OpticStudio design of the ELT where the seeing jitter (Figure \ref{ima_tip_tilt}) on point-like sources at infinity is introduced with a multi-configuration approach. The field is steered using the M5 tip-tilt degree of freedom calculated with a standard optimization based on a default merit function and some target operands to re-align the field against the atmospheric jitter.

\begin{table}
\begin{center}
\begin{tabular}{p{1.3cm}*{5}{c}} \hline
\textbf{Angle} &\textbf{FR(M4)} &  \textbf{FR(M5)} &  \textbf{Jitter(M4)} &  \textbf{Jitter(M5)} \\
\hline 
\bfseries $\theta_x=0.01^\circ$ & 9.7" & 43.6" & 1.84 mas & 8.28 mas\\
\bfseries $\theta_y=0.01^\circ$ & - & - & - & -\\
\bfseries $\theta_z=0.01^\circ$ & 10" & 24" &1.90 mas & 3.56 mas\\
\hline
\bfseries $\theta_x(seeing)$ & - & 12.64" & - & 2.4 mas\\
\hline
MICADO \\
\hline
\textbf{Field} & \textbf{PSF H} &  \textbf{PSF K}\\
(0",0") & 8.5 mas &10 mas\\
(0",30") & 10 mas & 13 mas\\
\end{tabular}
\caption{Top: Field Rotation (FR) and PSF jitter induced at the edge of 1 arcmin FoV by tilts of $0.01^\circ$ on M4 and M5 obtained with Eq. \ref{eq_matrices} \& \ref{eq_mirrormat}; Centre: FR and PSF jitter induced by typical tip-tilt corrections of the AO system by M5 for 1" seeing (tip-tilt values Figure \ref{ima_tip_tilt}); Bottom: typical PSF size of MICADO in H and K band. The combination of opto-mechanical positioning errors and tip-tilt correction can give origin to a FR that is a significant fraction / comparable size of the instrument PSF, smearing out the image at the edge of the FoV. \label{tab_FR}}
 \end{center}
 \end{table}
 
\begin{figure}
\centering\includegraphics[width=0.7\linewidth]{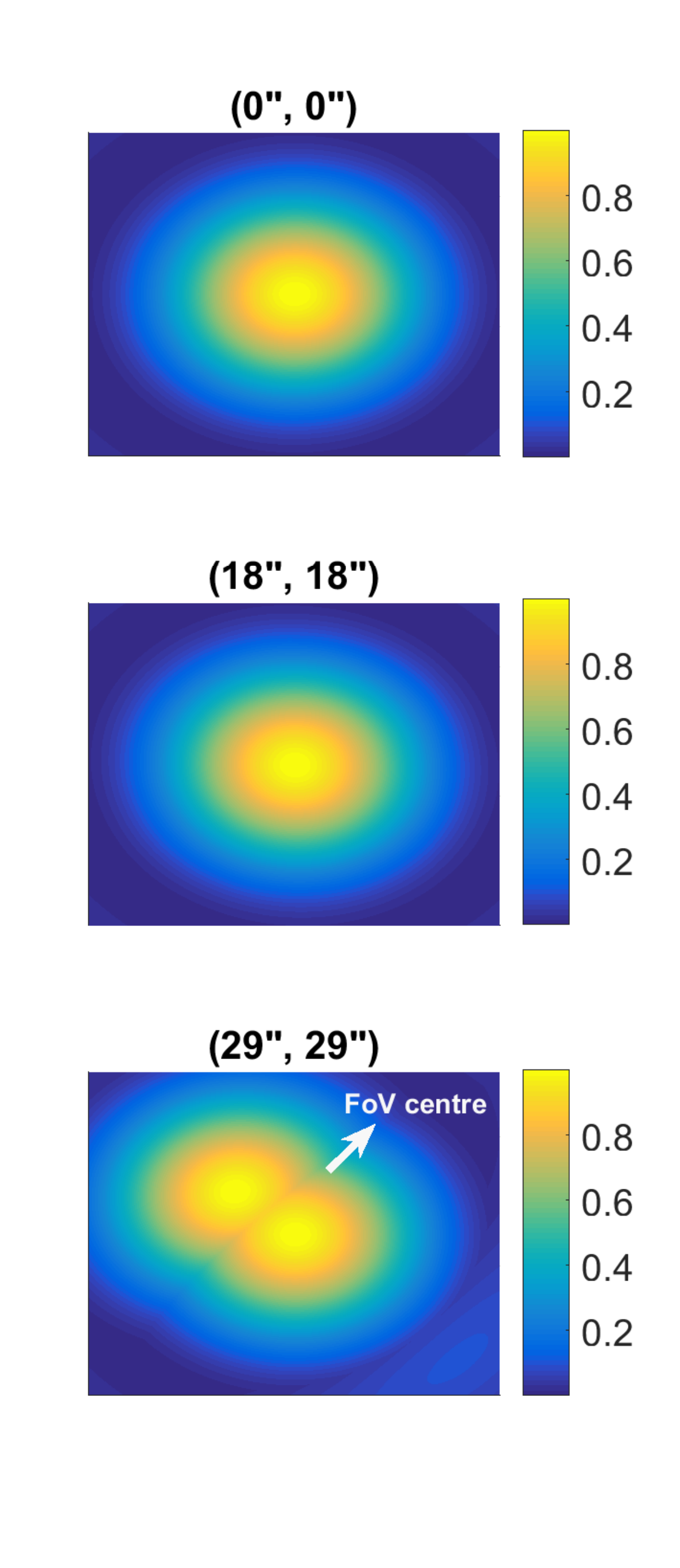}
\caption{Over plot of the nominal and 20 MC non-sequential ray tracing of the ELT PSFs at three different locations within 1 arcmin FoV.  The tip-tilt AO correction against the 1" seeing (Fig. \ref{ima_tip_tilt}) induces increasingly larger PSF jitter along the FoV due to the FR. While the on-axis and mid-FoV PSFs are aligned between different MC realizations, the off-axis PSFs are no longer piled up and they move along an arc whose center points to the telescope FP centre. The size of each detector is 60 $\mu$m. \label{fr_end}}
\end{figure}

The results of the ray tracing simulations are shown in Figure \ref{fr_end}: three small ideal detectors image the PSF at the ELT FP in three different field positions within 1 arcmin FoV. For a perfect alignment of the M5 (no FR) all the PSFs from different fields are centered on the detectors, while for the 20 random seeing realizations the PSFs of the off-axis fields distribute along an arc whose center points in the direction of the FP centre and whose length is proportional to the field position ($\sim$ 3.4 mas at (29", 29")). The amplitude of the PSF jitter at the FP is comparable with the number estimated with the matrix formalism (Eq. \ref{eq_mirrormat} \& \ref{eq_matrices}) and reported in Table \ref{tab_FR}. The FR induced by random or systematic tilts of M4 and M5 creates a PSF jitter that are a considerable fraction of typical instruments PSFs like e.g. MICADO (8.5-10 mas, $H$ band). Although in principle the slow opto-mechanical positioning errors could be tracked and compensated with the instrument derotation systems, the FR originated from the AO tip-tilt correction cannot be avoided thus posing some limitations to the accuracy of the centroiding of the PSF in the outer parts of the FoV. The combination of the PSF smearing due to FR and the intrinsically larger optical aberrations in the outer regions of the FoV leads to a degradation of the astrometric performances. 
An exhaustive estimate of the overall astrometric error would require taking into account also the PSF aberrations that are beyond the goal of this work, but the results of these assessment simulations pose a caveat to the error budget of the observations for relatively large fields.

\section{ELT on-sky calibration}

\begin{figure}
\centering\includegraphics[width=1\linewidth]{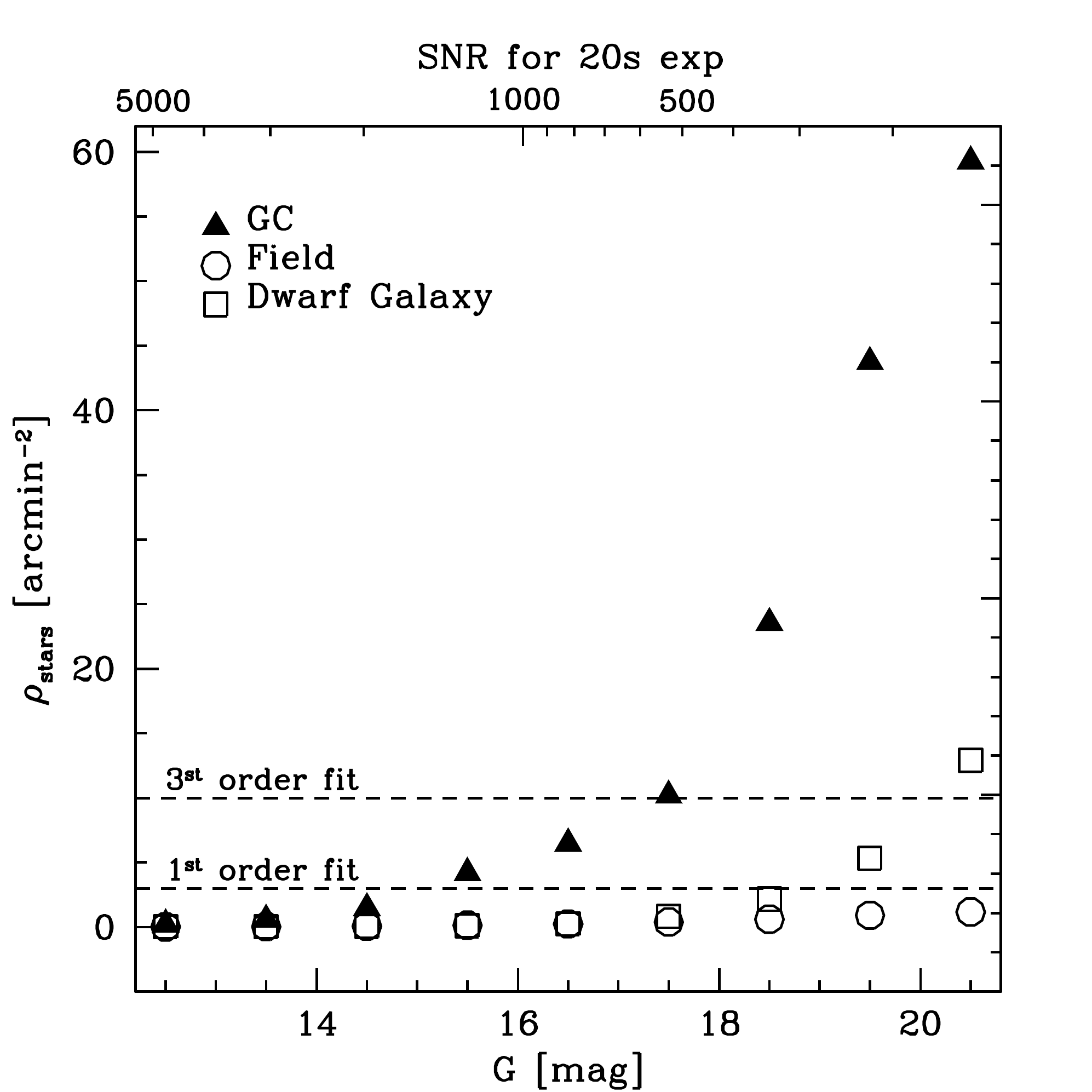}
\caption{{\bf Cumulative distribution of the} density of {\it Gaia} stars as function of magnitude for three environments with different crowding properties: a sparse stellar field (empty circles), a loosely crowded dwarf galaxy (empty squares) and a crowded globular cluster (filled triangles). The requirements to be able to calibrate first and third order distortions are shown as dashed lines. The expected SNR from 20 seconds long ELT observations is also quoted in the upper X-axis. The plot extends up to G = 21 mag in correspondence of the peak of completeness of the {\it Gaia} catalog.  \label{gaia_stars}}
\end{figure}

The currently foreseen strategy to correct the geometric distortions in astrometric observations with ELTs, and especially the low (up to third) order terms, heavily relies on on-sky calibrations.
There are two ways to do so. The first is to follow the self-calibration method developed e.g. by \cite{anderson2000}, which exploits the repeated observation of the same field with a proper dither strategy to model the distortion field affecting the camera. The second is to use astrometric standard stars, an approach which requires the a priori knowledge of the positions and proper motions of such standard stars, but which has already been successfully applied to MCAO observations by e.g. \cite{massari16}.
The latter method has the clear advantage of being much less demanding in terms of telescope time and temporal stability of the distortions, and for this reason, it will be the preferred channel for future instruments geometric distortion calibrations.
Standard stars to be used as reference will be mostly provided by astrometric missions such as {\it Gaia} \citep{prusti16} and Euclid \citep{Laureijs12}.  Since {\it Gaia} has recently provided new measurements in the Data Release 2 \citep{brown18, helmi18}, we are now able to perform an exploratory investigation on how many {\it Gaia} standard stars will typically be available over 1 arcmin$^2$ FoV to be used for distortion calibration. Figure \ref{gaia_stars} shows the cumulative density distribution as a function of magnitude of {\it Gaia} stars in three different environments: a sparsely populated Galactic field (empty circles),  a Galactic globular cluster (filled triangles) and an external dwarf spheroidal galaxy (empty squares).

\begin{figure}
\centering\includegraphics[width=1.05\linewidth]{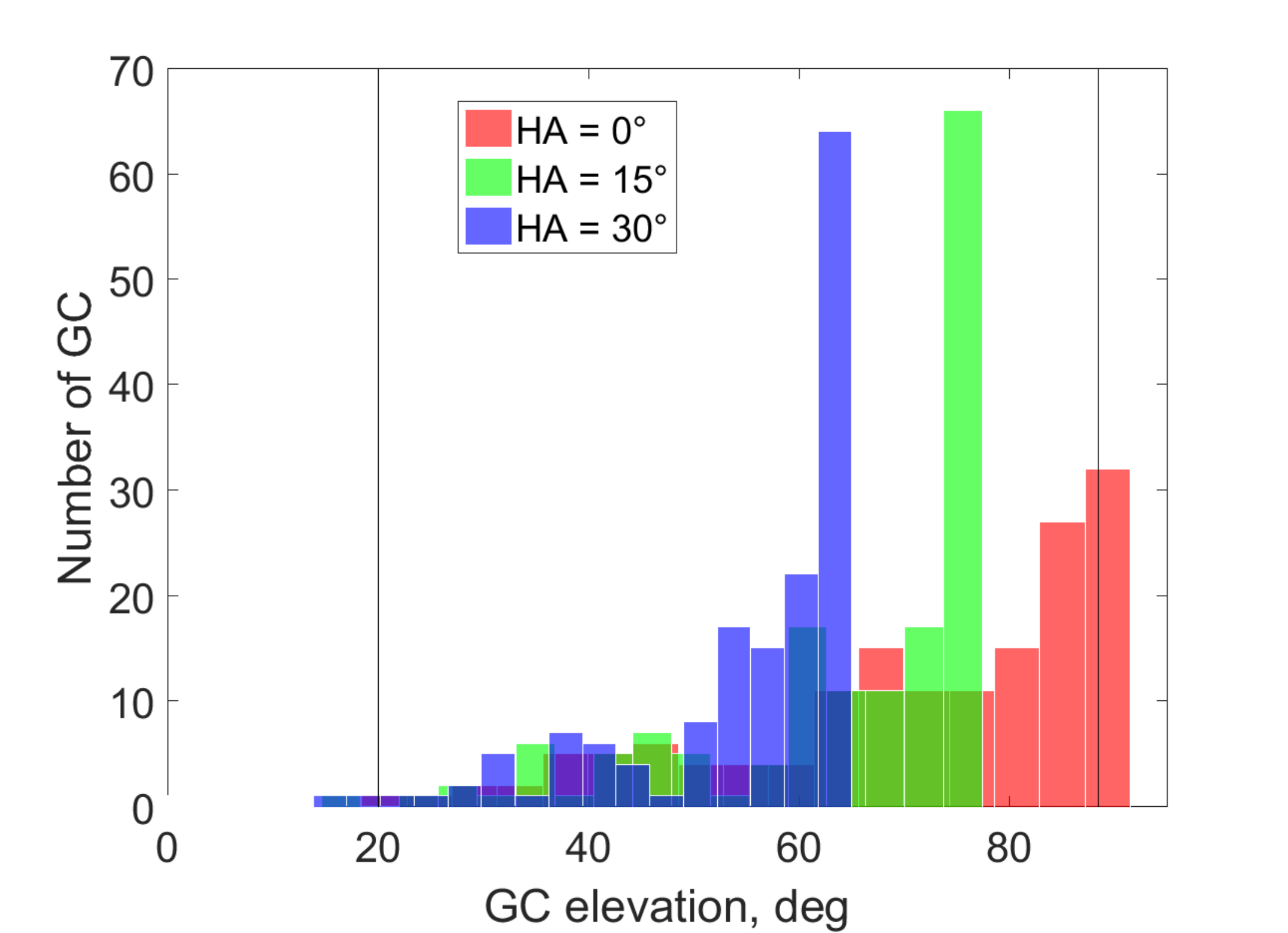}
\caption{Globular Clusters (GC) elevation at ELT latitude (-24$^\circ$ 40') for different hour angle (HA). A large number of GCs distributed in elevation is available for on-sky astrometric calibrations. The black vertical lines delimit the elevation range allowed to the telescope during science observations. \label{GC_elev}}
\end{figure}

It is clear that only in the crowded regions of a globular cluster the number of Gaia stars will be largely sufficient for our purposes: this is already true for the correction of linear terms (about 3 stars/arcmin$^{2}$ required, see the lower dashed line), while to correct third order distortions ($\sim$10 stars/arcmin$^{2}$ required) we will have to rely on stars as faint as G$\sim$ 18 mag (see the upper dashed line). Fainter stars (G$\sim20$) can also be used in the intermediately crowded field of dwarf spheroidal galaxies to correct third order terms. Whether the Gaia proper motions and position uncertainties will be good enough at these magnitudes, is beyond the scope of this paper and will be tested in future simulations. The proper motion accuracy of the \textit{Gaia} stars for G = 18 mag and 21 mag is expected to be $\sim$72.5 $\mu$as and $\sim$ 786.5 $\mu$as respectively \citep{deBruijne05}. To counterbalance the higher astrometric uncertainties of the fainter stars, one needs to calibrate the distortions over a larger number of stars $N$ so the uncertainties scale with $\sqrt{N}$.
In general, we can expect the feasibility of the astrometric science cases to be always guaranteed in globular clusters, and to be case dependent in Local Group dwarf galaxies, though the correction of linear terms seems always within reach. Where not feasible with {\it Gaia}, the correction of high order terms will be addressed by means of self-calibration.
We underline that what is faint for {\it Gaia} will be typically very bright for the future generation of ELTs instruments: the upper X-axis of Fig.\ref{gaia_stars} shows the expected SNR for a ELT exposure of 20 seconds (ELT ETC \cite{Liske2017}), which is larger than 100 for all the sources detected by {\it Gaia}. Moreover, the completeness of {\it Gaia} will improve, especially at the faint limit, as the survey goes on, so that the density estimate shown here can be effectively taken as a lower limit. To conclude the assessment study about the calibration of the ELT, we plot in Figure \ref{GC_elev} the elevation of a collection of Globular Clusters (GC) \citep{Harris1996} at the ELT latitude: a suitable number of GCs along the full elevation range accessible to ELT are available for on-sky astrometric calibrations, leading both to perform ad-hoc on-sky calibrations during science observations and to conduct dedicated, systematic studies of the distortions at different telescope pointing directions.

For the specific observation of the black hole in the galactic center, an astrometric reference frame can be tied to a set of red giant stars \cite{Plewa2015}.
\cite{Reid2007} reports the position of 15 red giant stars within 50 arcsec of Sgr A* emitting both at NIR and at radio wavelengths for their circumstellar SiO masers. The objects are measured with an accuracy of $\sim$1 mas in position and $\sim$0.3 mas yr$^{-1}$ in proper motion and within a FoV of $\sim$1 arcmin eleven sources with $m_H<$ 20 are available to calibrate the telescope distortions up to 3$^{rd}$ order (Table \ref{tab_astro}).

\begin{figure}
\centering\includegraphics[width=1.1\linewidth]{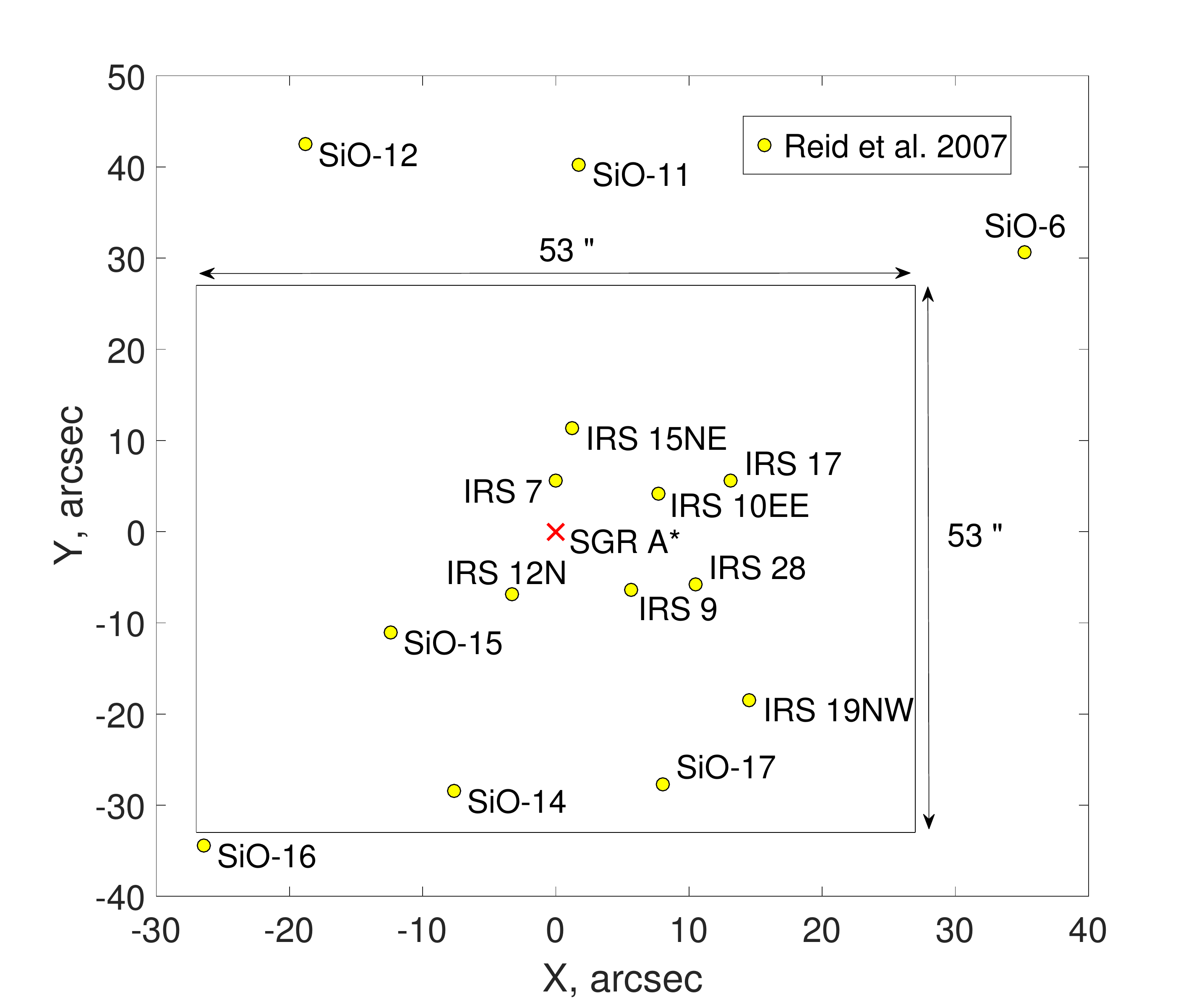}
\caption{SiO masers \citep{Reid2007} within $\sim$1 arcmin FoV available for the telescope distortion calibration on-sky up to 3$^{rd}$ order, $m_H<$ 20. \label{ima_masers}}
\end{figure}

\section{Conclusions}

We have modeled the telescope astrometric performances in a MonteCarlo approach by giving 3-D position and rotation errors in the tolerance range of $\pm$ 0.1 mm and $\pm$ 0.01$^\circ$. The figures of merit we have evaluated in our study are three: plate scale variation, exit pupil motion and RMS distortion over 1 arcmin FoV.
Comparing the telescope performance of each MC realization against the nominal design we find a maximum RMS distortion drift of $\sim$ 5 mas/arcmin, which is dominated by plate scale (M2-$\Delta$z motion) and 3$^{rd}$ terms, in good agreement with the expected PSF jitter reported in the ELT ICD. The ranges of tolerance assumed in our sensitivity analysis somewhat underestimate (factor $\sim$ 5) the expected exit pupil rotation and lateral displacement that might come from an underestimation of the M3-M5 positioning errors or from a global telescope motion/rotation of the main structure. The combination of the tolerances from all the ELT optics leads to low 1$^{st}$-3$^{rd}$ distortion terms that reduce respectively to $\sigma_{post-1st} \sim 20-100$ $\mu as$ and $\sigma_{post-3rd} \sim 10-20$ $\mu as$ after the polynomial fit. 
The distortion drift between two consecutive LOO loops ($\sim$ 5 min) is driven by the M2 that produces mostly PS variations whose impact on astrometric observations can be controlled with an MCAO system.
Preliminary opto-mechanical design analysis of the ELT suggest that  wind shakes predominantly move M2 at 0.1-2 Hz. Variation of the M2 position by $\Delta z \sim 0.25 $ mm leads to a PS rms drift $\sigma_{OL}(\Delta PS) \sim 18$ $\mu$as/mm that can be suppressed significantly by two DMs operated in a MCAO system. 
This order of magnitude of wind shake impact on the M2 position is consistent with the preliminary dataset of kinematic analysis of the ELT released to the instrument Consortia (\cite{Marchetti2015}, \cite{Schmid2017}). No significant skewness in the PS along the field has been pointed out by the simulations.
The random field rotation induced by random or systematic tilts of M4 and M5 creates a PSF jitter during an exposure increasing with the radius of the field. Unavoidable AO tip-tilt seeing correction induces a PSF jitter at the edge of 1 arcmin FoV that is a significant fraction ($\sim$ 2-2.5 mas) of typical instruments PSFs for ELTs.
An important topic to be refined for future studies is the rate of change of the telescope optics positioning errors. The current study gives, for assumed values, the extreme RMS estimates, but ultimate reference values would address more precisely the instruments calibration plans and the calibration overhead on-sky in typical observation scenarios.

\section{Acknowledgments}

We are sincerely grateful to the anonymous referee for his review work. We warmly acknowledge for the useful discussions and comments Dr. Enrico Marchetti, Steffan Lewis, Claudio Pernechele, Conchi C\'{a}rdenas V\'{a}zquez, Stefan Gillessen, Ralf Rainer Rohloff, Enrico Pinna.











\bsp	
\label{lastpage}
\end{document}